\documentclass[fleqn,usenatbib]{mnras}

\usepackage{newtxtext,newtxmath}

\usepackage[T1]{fontenc}
\usepackage{ae,aecompl}

\usepackage{graphicx}	
\usepackage{amsmath}	
\usepackage{amssymb}	
\usepackage{multirow}
\usepackage{float}
\usepackage{caption}
\makeatletter
\def\@to{to}
\makeatother

\newcommand{\hckpc}{\:\mathrm{h^{-1}ckpc}}
\newcommand{\hpkpc}{\:\mathrm{h^{-1}pkpc}}
\newcommand{\si}{\mathrm{Si}}
\newcommand{\siii}{\mathrm{Si}\:\textsc{ii}}
\newcommand{\siiv}{\mathrm{Si}\:\textsc{iv}}

\newcommand{\civ}{\mathrm{C}\:\textsc{iv}}

\newcommand{\oxygen}{{\mathrm{O}}}
\newcommand{\oi}{{\mathrm{O}\:\textsc{i}}}

\newcommand{\ov}{{\mathrm{O}\:\textsc{v}}}
\newcommand{\ovi}{{\mathrm{O}\:\textsc{vi}}}

\newcommand{\hi}{\mathrm{H}\:\textsc{i}}
\newcommand{\hii}{\mathrm{H}\:\textsc{ii}}
\newcommand{\hydrogen}{\mathrm{H}}

\newcommand{\hei}{\mathrm{He}\:\textsc{i}}
\newcommand{\heii}{\mathrm{He}\:\textsc{ii}}
\newcommand{\angstrom}{\mbox{\,\normalfont\AA}}

\title[Neutral Oxygen During Hydrogen Reionization]{Evolution of Neutral Oxygen During the Epoch of Reionization and its Use in Estimating the Neutral Hydrogen Fraction}

\author[C. Doughty et al.]{
Caitlin Doughty,$^{1}$\thanks{E-mail: cdoughty@nmsu.edu}
Kristian Finlator,$^{1}$
\\
$^{1}$New Mexico State University, Las Cruces, New Mexico 88001\\
}

\date{Accepted XXX. Received YYY; in original form ZZZ}

\pubyear{2019}
\begin{document}
\label{firstpage}
\pagerange{\pageref{firstpage}--\pageref{lastpage}}
\maketitle

\begin{abstract}
We use synthetic sightlines drawn through snapshots of the Technicolor Dawn simulations to explore how the statistics of neutral oxygen ($\oi$) absorbers respond to hydrogen reionization. The ionization state of the circumgalactic medium (CGM) initially roughly tracks that of the intergalactic medium, but beginning at $z=8$ the CGM grows systematically more neutral owing to self-shielding. Weak absorbers trace diffuse gas that lies farther from halos, hence they are ionized first, whereas stronger systems are less sensitive to reionization. The overall $\oi$ covering fraction decreases slowly with time owing to competition between ongoing enrichment and gradual encroachment of ionization fronts into increasingly overdense gas. While the declining covering fraction is partially offset by continued formation of new halos, the ionization of the diffuse gas causes the predicted line-of-sight incidence rate of $\oi$ absorbers to decline abruptly at the overlap epoch, in qualitative agreement with observations.  In comparison to the recently-observed equivalent width (EW) distribution at $z\approx6$, the simulations underproduce systems with $\mathrm{EW} \geq 0.1 \angstrom$, although they reproduce weaker systems with $\mathrm{EW} \geq 0.05 \angstrom$. By $z\approx5$, the incidence of $\mathrm{EW} < 0.1 \angstrom$ systems are overproduced, consistent with previous indications that the simulated ionizing background is too weak at $z<6$. The summed column densities of $\siii$ and $\siiv$ trace the total oxygen column, and hence the ratio of the $\oi$ and $\siii + \siiv$ comoving mass densities traces the progress of reionization. This probe may prove particularly useful in the regime where $x_{\hi} > 10\%$.
\end{abstract}




\section{Introduction}\label{sec:introduction}
The epoch of reionization (EoR) saw the transition of hydrogen from a neutral state to an ionized state. From $z=30\rightarrow10$, early structures began to form~\citep{couchman86, haiman97, ostriker96} and Population III stars began to contribute hard ionizing photons to the process of hydrogen reionization. However, the EoR didn't see drastic changes in the neutral hydrogen fraction until larger structures, like galaxies and galaxy clusters, were well-established. It is generally believed that reionization was underway by $z\sim13$, more than halfway finished by $z\sim7$, and completed by $z\sim6$, although significant questions remain regarding the timing and end-stages~\citep{fan06, glazer18, hoag19, morales10, inoue18, mason18, planckcol18, ouchi18}. The duration, timing, and geometry of reionization are dependent on the spectral characteristics and spatial distribution of the ionizing sources. Lyman-$\alpha$ (Ly$\alpha$) forest observations at $z>5$ suggest significant spatial variability in the opacity of the intergalactic medium (IGM) that could be explained by the presence of rare but luminous objects such as Active Galactic Nuclei (AGN)~\citep{chardin15,chardin17,daloisio17}. However, AGN influence the state of the post-hydrogen-reionization IGM, contributing to premature $\heii$ reionization and elevated IGM temperatures that conflict with observations~\citep{daloisio17}. Further, study of the quasar luminosity function at high-redshifts suggests they may not be abundant enough to sustain reionization~\citep{onoue17,mcgreer18}.

Regardless of the exact nature of the sources of ionizing photons, the expansion of ionization fronts around the sources resulted in a ``patchy'' universe containing regions with high neutral fractions of hydrogen with corresponding high optical depths alongside regions with high flux transmission~\citep{mcquinn07}. This theoretical picture is supported by the observed sightline-to-sightline scatter in the mean transmission of the Ly$\alpha$ forest at $z>5.5$, which is too large to reflect density fluctuations alone~\citep{becker15mnras}.

There are many complementary probes of hydrogen reionization. The most direct approach is to examine quasar spectra for intervening Ly$\alpha$ absorption, a sign of significant neutral hydrogen along the line of sight. However, this tool is limited by the ease with which Ly$\alpha$ absorption saturates: for measurable transmission, the volume-averaged neutral hydrogen fraction must be less than $10^{-3}$, otherwise the signature becomes saturated and difficult to use in distinguishing between different $\hi$ quantities~\citep{fan02}. Consequently, this probe is useful only for investigation of the very latest stages of reionization, when the neutral fraction was already quite low. For very high neutral fractions, the Gunn-Peterson trough~\citep{gunn65} observed in a quasar spectrum may be examined for a redward damping wing, but the hard ionizing radiation surrounding quasars is unfortunately also capable of obscuring this feature, preventing its use (see e.g.~\citet{miralda98, madau00}).
It is also possible to examine the progression of reionization by analyzing the number density and lengths of the Gunn-Peterson troughs between transmission spikes in Ly$\alpha$ at very high redshifts. However, the use of such higher-order statistics is not yet widespread~\citep{mortlock16}.

Ly$\alpha$ emitters (LAEs) are also sensitive to reionization. The observed Ly$\alpha$ luminosity function (LF), paired with simulations, has been used to determine the ionized fraction of hydrogen based on the decrement between simulated and observed LFs, or by examining the redshift evolution of the LF for $z>5.7$~\citep{inoue18,konno18}. Similarly, examining the deviation of observed LAE clustering from their theoretical intrinsic correlation function may indicate the fraction of Ly$\alpha$ sources which are obscured by neutral hydrogen~\citep{ouchi18}. However, these methods hinge on the assumption that Ly$\alpha$ physical characteristics and populations do not evolve significantly during the EoR, which \text{is} not as-yet verified. Indeed, some observations have suggested that there may be a dropoff in the number of LAEs at $z>6$~\citep{treu13}. Further, galaxies bright in emission are predisposed to exist in overdense regions, meaning that they are not necessarily representative of all structure during reionization (see e.g.~\citet{einasto87}).

Redshifted 21 cm emission can provide another avenue of understanding reionization. For example, the all-sky 21 cm signal would provide information about the formation of the first stars at cosmic dawn leading to the initiation of reionization, and indeed there has been a tentative detection~\citep{bowman18}. However, such observations are not yet available in bulk and are still difficult to obtain, in part due to issues with foreground contamination~\citep{patil17}. Forthcoming tomography in 21 cm $\hi$ emission, on the other hand, will allow direct observation early structures and galaxies as they form, as well as for the characterization of changes in the $\hi$ abundance~\citep{morales10}. These observations lie on the horizon, as facilities such as the Hydrogen Epoch of Reionization Array (HERA) are built~\citep{deboer16}. In addition, the Square Kilometer Array (SKA) will have a low-frequency component, the Low Frequency Array (LOFAR), devoted to 21 cm tomography~\citep{mellema13}.

Metal absorption arising in quasar absorption spectra can serve as an independent probe of reionization. Distinct from absorption coming from gas in the quasar's host galaxy, this metal absorption consists of signatures from the circumgalactic medium (CGM) of foreground galaxies. At low to intermediate redshifts, metal absorbers are used as probes of CGM gas kinematics and star formation activity~\citep{prochter06, nielsen17, tumlinson17}. Other works further suggest a relationship between absorbers and host halo properties, such as mass~\citep{oppenheimer16,kacprzak19}.  
There is also evidence of causal connection between galaxies and absorbers at very high redshifts ($z>5$) with characteristics of early galaxies, such as their masses or star formation rates, showing some correlations with absorber equivalent widths or galaxy-absorber separation~\citep{cai17,garcia17}. Though the quasar number density drops significantly above $z>4$, there are still a significant number of visible quasar lines-of-sight over the entire sky for $z>5.6$~\citep{banados16}. It is also possible to use the afterglow of very high redshift gamma ray bursts to glimpse absorption in the CGM and IGM, enhancing the usefulness of predictions for high redshift metal absorbers~\citep{stark16}.

In this work we focus on $\oi$, which has been proposed as a tracer of hydrogen reionization~\citep{oh02,finlator13,keating14}. Because of the similarity of the $\oi$ ionization energy to that of neutral hydrogen, it is sensitive to the same photons that drive hydrogen reionization. Further, because of the relatively low column densities of oxygen compared to hydrogen, $\oi$ absorbers are less likely to be saturated. Consequently, they may constrain the evolution in the neutral hydrogen fraction induced by the rapidly-evolving ultraviolet background (UVB).  A theoretical $\oi\;\lambda 1302 \angstrom$ ``forest'', analogous to the Ly$\alpha$ forest, may even be visible in high redshift quasars, if there are no other spectral features to obscure it~\citep{oh02}.

The central theoretical question that must be addressed in order to understand how $\oi$ responds to reionization is: are $\oi$ absorbers expected to trace gas that is predominantly neutral or ionized on large scales? For the hypothetical case of a constant, spatially-uniform metallicity, it is obvious that $\oi$ traces either dense or un-reionized regions. In reality, however, $\oi$ must lie near to the galaxies where oxygen is created. If galaxies dominate ionizing photon production, then $\oi$ absorbers trace regions where, on large scales, the UVB is stronger than the cosmic average. Simulations have shown generally that low-ionization states similar to $\oi$ show a tendency to lie in the inner CGM ($r < 0.5 R_{200}$), at least for lower redshifts, while higher ionization states are in the diffuse IGM~\citep{oppenheimer18}.

This question illustrates how accurate modeling of the relationship between $\oi$ and reionization is a complicated endeavor: First, models must capture the growth of ionized regions, which can be $\sim10$ Mpc by $z=6$. Next, they must model small-scale hydrodynamical processes such as galactic outflows and their interaction with the CGM. These are treated only approximately in cosmological simulations, which may lead models to underpredict strong systems~\citep{keating16}. Finally, faithful application of an inhomogeneous ionizing background to gas particles is required to reduce conflict with observed $\oi$ abundances. A spatially-invariant (absent) extragalactic ultraviolet background has been shown to lead to underproduce (overproduce) $\oi$ in cosmological previous simulations~\citep{oppenheimer09}. It is perhaps for these reasons that the use of $\oi$ as a probe of reionization remains in its infancy.

Nonetheless, there are hints of evolution in $\oi$ suggestive of sensitivity to reionization. Although there has been no definitive detection of a forest in $\oi$, it has been observed that the abundance of low-ionization absorbers, including $\oi$, increases with redshift at $z\geq5.7$. \citet{becker11} identified 10 low-ionization absorption systems including $\oi$ along 17 quasar lines-of-sight at $5.3<z<6.4$, with all systems located at $z>5.7$. Their expanded investigation in~\citet{becker19} was able to detect 57 non-proximate $\oi$ systems in 199 quasar spectra across $3.2 < z < 6.5$, with the line-of-sight number density increasing between two bins divided at $z=5.7$. This increase, in conjunction with the decrease at $z\geq5.3$ of high-ionization systems such as $\civ$, is further evidence of the changing UVB in this redshift regime~\citep{ryan-weber09}. Observing four high-redshift quasars ($5.79 \leq z_\mathrm{em} \leq 6.13$), \citet{codoreanu18} detect no $\oi$ systems, perhaps either confirming the end of reionization or suggesting a high degree of variability in the number of $\oi$ systems visible on a given sightline. The rapid decline in the incidence of $\oi$ systems, if confirmed through future observations, may stem primarily from abrupt changes in the ionizing background rather than evolution in the enrichment or density of the IGM or CGM~\citep{keating14}, although in practice all of these effects contribute at some level~\citep{finlator15}.

In summary, theoretical studies suggest that $\oi$ is responsive to reionization and there are observational hints that this may already have been observed. In order to understand these observations, we undertake a new inquiry into how $\oi$ responds to $\hi$ reionization that improves on previous theoretical study in several ways. First, our radiation-hydrodynamic simulation is significantly more realistic because it spans a larger dynamic range with a multifrequency radiation transport solver. Second, the simulation is calibrated to reproduce an array of complementary observations of reionization~\citep{finlator18}. Finally, whereas~\citet{finlator13} measured the predicted $\oi$ covering fraction as a function of halo mass and then computed the resulting absorber statistics from the dark matter halo mass function, we now create realistic synthetic sightlines and compile absorbers in a way that closely mimics observations. In this work, we use this improved theoretical framework to address three related questions regarding  neutral oxygen during the EoR:
\begin{enumerate}
    \item How do the neutral oxygen halos of galaxies evolve with time?
    \item Is neutral oxygen found predominantly in enriched regions of the IGM that have not yet been reionized, or in dense, self-shielded portions of the CGM?
    \item Can the statistics of OI absorbers be used to trace the progress of hydrogen reionization?
\end{enumerate}
In Section~\ref{sec:simulations} we discuss the relevant details of the simulation. In Section~\ref{sec:results} we present the evolution of the properties of the absorber population, the changing spatial distribution of $\oi$, and the relationship between $\oi$ and halos as well as with neutral hydrogen. In Section~\ref{sec:discussion} we discuss the implications of our results and place them in context given other work. We then present our conclusions in Section~\ref{sec:conclusions}. Unless otherwise noted, we assume a $\Lambda \mathrm{CDM}$ cosmology with $\left(\Omega_M, \Omega_{\Lambda}, \Omega_b, h, X_H\right) = \left( 0.3089, \; 0.6911, \; 0.0486, \; 0.6774, \; 0.751\right)$ in our calculations~\citep{planckcol16}.

\begin{figure*}
\centering
\includegraphics[width=1.0\textwidth]{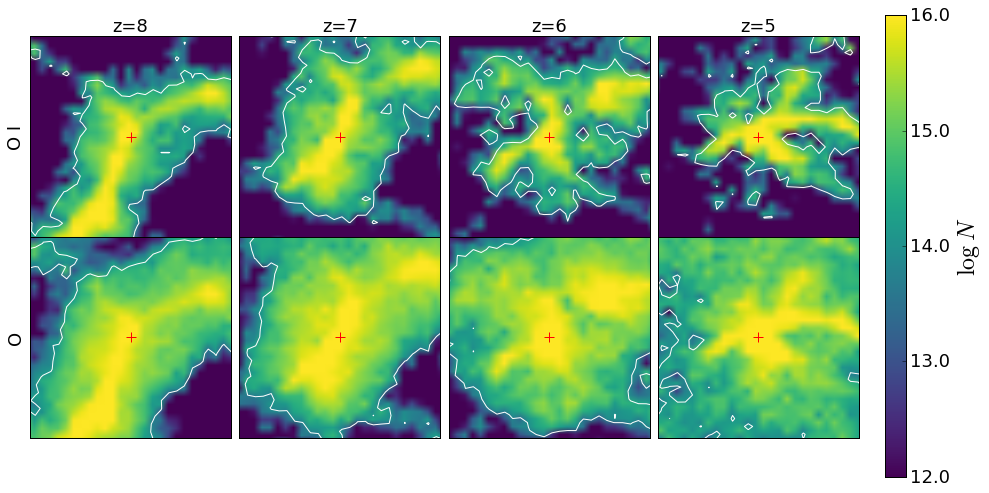}
\caption{Slice of width $\delta v = 50$ km/s centered on the most massive halo, defined at $z=5$ to be $M_\mathrm{halo} = 1.1\times10^{11} M_{\odot}$. Color corresponds to column density, of $\oi$ in the top row and total oxygen in the bottom row. The red cross indicates the location of the halo's center at each redshift. The white line shows the location of the $\log\;N = 14.0$ contour. Imaged box side length corresponds to $0.6 \; \mathrm{h}^{-1} \mathrm{cMpc}$.} \label{fig:OI_O_coldens_map}
\end{figure*}

\section{Simulations}\label{sec:simulations}
\subsection{Cosmological simulations}\label{ssec:cosmosims}
We draw our predictions from the fiducial ``n512RT64'' Technicolor Dawn simulation previously presented in~\citet{finlator18}. We summarize the model here and refer the reader to that paper for more details. We model galaxy growth and reionization within a $12\;h^{-1}\;\mathrm{Mpc}$ volume. Matter is discretized into $2\times512^3$ dark matter and gas particles. The simulation assumes a Planck cosmology~\citep{planckcol16} with $\left(\Omega_M, \Omega_{\Lambda}, \Omega_b, h, X_H\right) = \left( 0.3089, \; 0.6911, \; 0.0486, \; 0.6774, \;0.751 \right)$. Initial conditions are generated using {\sc music}~\citep{hahn11}.

Dense gas acquires a subgrid multiphase structure and forms stars following~\citet{springel03} at a rate that is regulated primarily by galactic outflows. The outflow mass-loading factor is given by the stellar mass-dependent scaling in Equation 8 of~\citet{muratov15}, while wind velocity is calculated using a slightly modified version of the methods in~\citet{dave16}. Metal enrichment is tracked in ten independent species and receives contributions from asymptotic giant branch stars as well as Type Ia and Type II supernovae (SNe). SNe yields are computed from the~\citet{nomoto06} yield tables assuming a~\citet{kroupa01} initial mass function and a 50 percent hypernova fraction. The rate assumes a time delay distribution taken from~\citet{heringer17}. Metal enrichment owing to asymptotic giant branch stars is accounted for following~\citet{oppenheimer08}.

Our Technicolor Dawn simulation couples these treatments for galaxy growth and feedback with an on-the-fly radiation transport solver in order to model the growth of a spatially inhomogeneous UVB and the progress of reionization self-consistently, using the full speed of light. We discretize the moments of the radiation field into 24 independent frequency bins spaced evenly in energy from 1-10 Rydbergs and spatially onto a regular grid of $64^3$ volume elements (``voxels''). Each voxel's emissivity is given by a metallicity-weighted sum over the star formation rates of its gas particles, and its opacity owes to bound-free transitions of $\hi$, $\hei$, and $\heii$. We evolve the UVB in time using a moment method~\citep{finlator09,finlator15}.   As the voxel light-crossing time is large compared to the dynamical time within dense regions, no need arises for approximations regarding the speed of light. An iterative solver guarantees that the cooling/ionization and radiation updates converge at each timestep.  Attenuation of the UVB in dense regions owing to ``self-shielding'' is accounted for via a subgrid treatment that assumes hydrostatic and ionization equilibrium; its predictions are in excellent agreement with the results of high-resolution radiation transport calculations~\citep{finlator15}.

Our simulation also tracks the quasar contribution to the UVB using a volume-averaged calculation. However, quasars are predicted to be subdominant, accounting for no more than a (6, 22) percent of $\hi$ photoionizations at $z>(6,\ 5)$. For this reason, we do not review it and refer once again to~\citet{finlator18} for details.

\begin{figure*}
\centering
\includegraphics[width=1.0\textwidth]{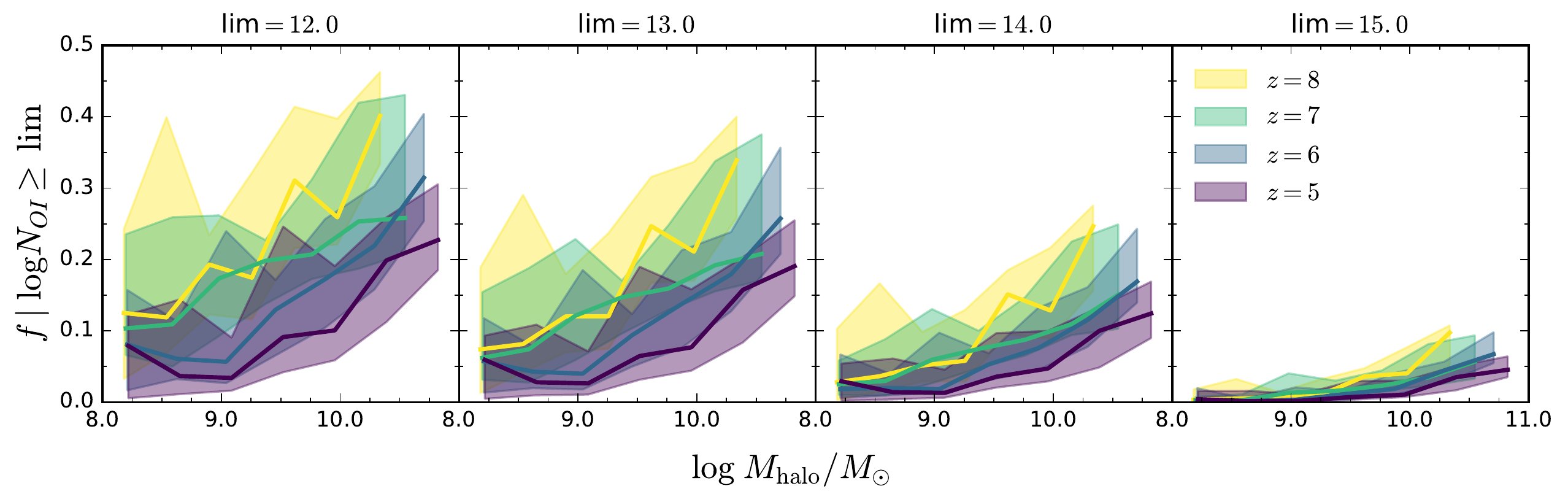}
\caption{$\oi$ covering fractions for different column density limits as a function of halo mass for 100 halos, calculated for an $r= 500 \; \hckpc$ circle centered on each halo. The regions between the 16th and 84th percentiles of the distributions are shaded to show the 1$\sigma$ variation with the median for each halo mass bin indicated by the solid line. \label{fig:covering_fraction}}
\end{figure*}

\subsection{Simulated Absorbers}\label{ssec:simulated_absorbers}
The procedure for creating the simulated spectra generated by metal absorbers follows closely the descriptions given in~\citet{finlator15} and~\citet{doughty18}, with a few exceptions. We begin by casting a sightline through the entire simulation box, oblique to the boundaries, that wraps around the box until it has covered a specified velocity width. The optical depth in each ion is calculated by accounting for contributions from particles encountered by the sightline. An absorber is identified where the transmitted flux drops $5\sigma$ below the continuum, and the velocity position in the sightline is output along with the column density, equivalent width, and several other line parameters. An artificial S/N of 50 is added to each resolution element of the spectrum.

From this output catalog, we merge together any absorbers of the same ion falling within 50 km/s of one another, summing the column densities and defining the column-weighted central velocity. The choice of 50 km/s is based on a commonly used convention in the literature (see e.g.~\citet{songaila01}). Testing the sensitivity of the chosen velocity cutoff, we find that there is an insignificant adjustment to the column density distribution of systems by increasing the value up to 200 km/. Regarding the deviations from the previous treatment, the most significant change is that in place of {\sc autovp}, the absorption features are created using a Voigt profile~\citep{humlicek79} (although this adjustment made no significant difference in the $\oi$ absorber statistics, since our absorbers are in the linear portion of the curve of growth). The final modification is a change in the sightline length, which corresponds to a Hubble velocity of $4\times10^6$ km/s, increasing the overall pathlength compared with that in the previous work. The velocity resolution in the sightline is 2 km/s.

\section{Results}\label{sec:results}
\subsection{Spatial distribution}\label{ssec:spatial_distribution}
To build intuition about the changes in the $\oi$ spatial distribution, we begin by extracting slices of a comoving volume containing the most massive halo (defined at $z=5$ with $M_\mathrm{halo} = 1.1\times10^{11} M_{\odot}$) at different redshifts.  We visualize the trends in this overdense region in both $\oi$ and the overall oxygen column density (top and bottoms rows, respectively, of Figure~\ref{fig:OI_O_coldens_map}). For this halo, it is apparent at a glance that the distribution of $\oi$ changes significantly in the $z=8\rightarrow5$ period. At $z=8$, there is a large swath of $\oi$ structures with column densities of $\log N_{\oi} \geq 16.0$, with a gradual decrease in column density as the distance from the central structure increases. Progressing from $z=8\rightarrow7$ there is an apparent increase in systems of $\log N_{\oi} \geq 15.0$ and 16.0 as the structure puffs up and encompasses more total area. Dropping to $z=6$, however, there is a recession of the $\oi$ boundaries as the highest column densities retreat into the innermost parts of the structure.  By $z=5$, the $\oi$ appears to be mostly grouped into very high or very low column densities with a sharp discontinuity separating the two populations. The expansion of the oxygen halo, visible in the bottom row of Figure~\ref{fig:OI_O_coldens_map}, is driven by transportation of metals into the CGM and IGM by simulated galactic winds. Its growth, coeval with the decrease in $\oi$ cross-section, demonstrates that the change in $\oi$ is not due to decreasing enrichment of the CGM, but may be stemming from an evolving UV background.

To establish the ubiquity of shrinking $\oi$ cross-sections from $z=8\rightarrow5$, we plot the covering fractions in $\oi$ for 100 halos in each redshift snapshot to see how they change for the overall population, where the halos were selected to sample the entire range of halo masses greater than $10^8 M_{\odot}$ in log space.\footnote{Due to the mass resolution of the simulation, halos below this mass will have artificially low gas densities and thus suppressed star formation rates.} The covering fractions are shown as a function of halo mass in Figure~\ref{fig:covering_fraction}. It is apparent that regardless of the exact column density limit, the covering fraction of $\oi$ generally decreases with decreasing redshift from $z=8\rightarrow5$ and increases with increasing halo mass. For any column density limit, there is significant overlap between the $z=8$ and 7 distributions (although the 84th percentile of $z=8$ does consistently reach higher values than $z=7$) and the $z=6$ and 5 distributions overlap significantly. Between $z=8$ and 7, the median covering fraction for all limits decreases more for halo masses greater than $\approx 10^{9.3} M_{\odot}$ than for smaller ones. Between $z=7$ and 6, the median covering fraction seems to be changing most for an intermediate range of halo masses, between the displayed minimum of $10^8 M_{\odot}$ and $10^{10.5} M_{\odot}$. In the interval between $z=6$ and 5, the largest changes appear to be occurring again for halos with masses over $10^9 M_{\odot}$. Another item of note is that while the slopes of the $z=8$ and 7 covering fraction distribution are \textit{very} roughly linear, for $z=6$ and 5 there is a knee in the relationship where below $M_\mathrm{halo} \leq 10^9 M_{\odot}$ the slope is almost flat and above $10^9 M_{\odot}$ increases sharply.

\subsection{Absorber Population Characteristics}\label{sec:population_characteristics}
In this section, we analyze the redshift evolution of $\oi$ systems. To begin, we define the absorption pathlength, $dX$, based on~\citet{bahcall69}, as
\begin{equation}\label{eq:std_pathlength}
dX(z) = \frac{H_0}{H(z)} \left(1 + z \right)^2 \; dz
\end{equation}
where $dz = \frac{dv}{c} \left(1+z\right)$, and $dv$ is the velocity spread of the sightline. We will refer to the number of absorbers per absorption pathlength, also called the incidence rate, as $\ell \left(X\right)$ for the remainder of this work. Since $\oi$ is a low-ionization state existing preferentially in dense gas, the expectation is that if evolution of the distribution is driven by the ionizing background, systems at lower EW in more diffuse gas will be ionized before higher EW systems.

\begin{figure}
\centering
\includegraphics[width=0.5\textwidth]{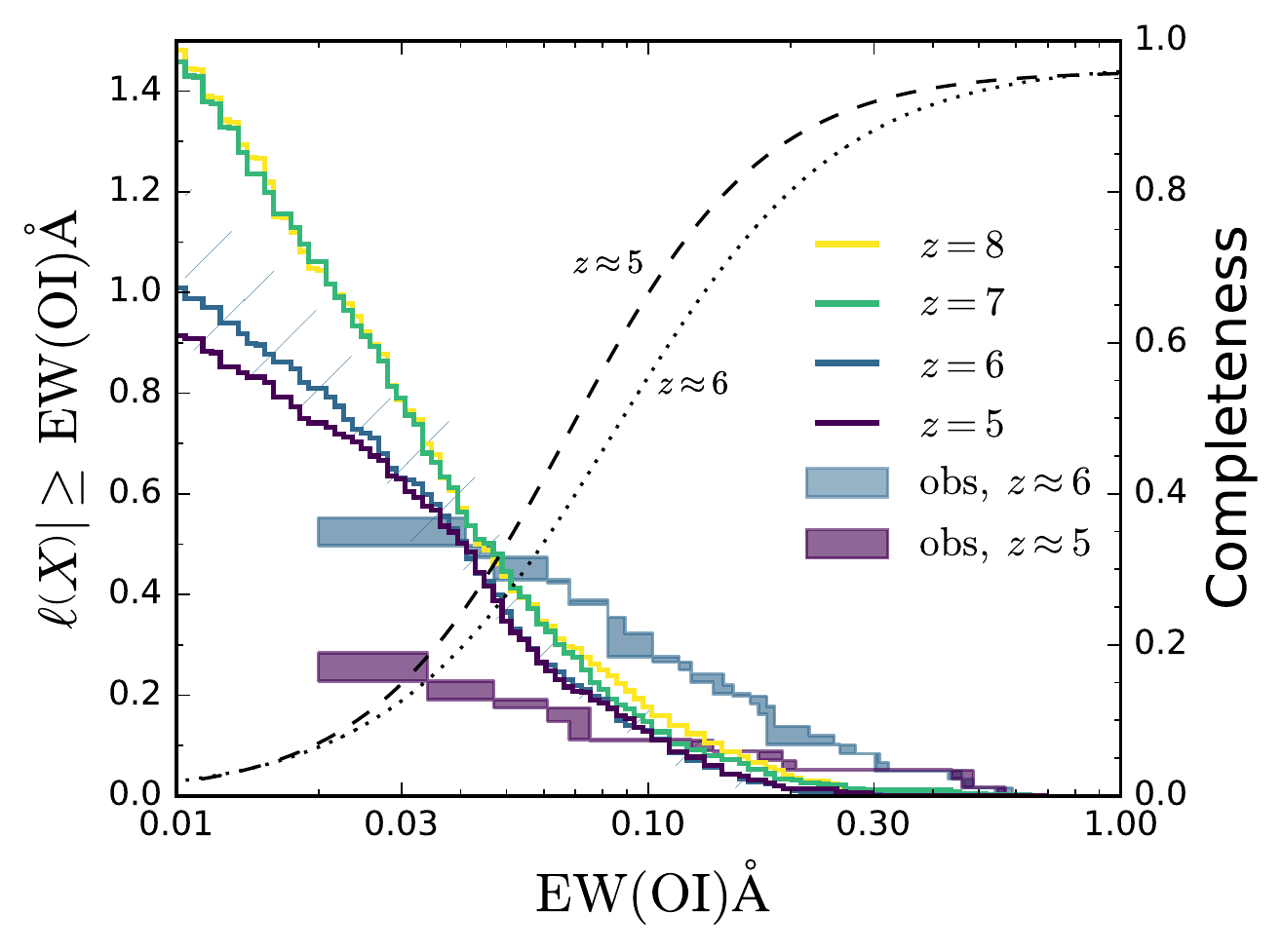}
\caption{Cumulative histogram of the distribution of $\oi$ equivalent widths from the simulation alongside the observations of~\citet{becker19}, all normalized by the pathlength at the given redshift as calculated using our cosmology (note the logarithmic scale on the x-axis). The blue hatching shows the scatter about the $z=6$ distribution when the long sightline is divided into segments of $dX=63.3$, comparable to that of the observations at $z=6$. The shaded regions show the 1$\sigma$ scatter around the observed EW values calculated by selecting from a random Gaussian distribution about the mean within the observational uncertainty for each measurement. A completeness correction is applied according to Equation~\ref{eq:compl_fit_function}, a fit of the curves in Figure 2 of~\citet{becker19}. The completeness for $4.9 < z < 5.7$ ($5.7 < z < 6.5$) is shown in a black dashed (dotted) curve.\label{fig:EW_OI}}
\end{figure}

We begin by examining the evolution of the cumulative equivalent width (EW) distribution from $z=8\rightarrow5$ (see Figure~\ref{fig:EW_OI}, and note the logarithmic scale on the x-axis). Serving as a point of comparison are data from \citet{becker19}. The impact of observational incompleteness is modeled by fitting the completeness curves of~\citet{becker19} for both their $4.9 < z < 5.7$ and $5.7 < z < 6.5$ observations assuming a sigmoid function of the form
\begin{equation}\label{eq:compl_fit_function}
f \left(x\right) = \frac{L}{1 + e^{-k \left(x - x_0\right)}}
\end{equation}
using notation as in~\citet{keating16}. For $z < 5.7$, the fit parameters are $\left(L, k, x_0\right) = \left(0.96, 4.81, -1.17\right)$, while for $z>5.7$ they are $\left(0.97, 4.20, -1.07\right)$.

From $z=8\rightarrow5$, each decrease in redshift shows a decrease in the total number of systems per pathlength, where at $z=8$ and 7 the cumulative EW distribution displays a steep slope that becomes flatter by $z=6$ and $5$. From $z=8\rightarrow7$ and $z=7\rightarrow 6$, there are minute decreases in the number density of $\oi$ absorption systems with $\mathrm{EW} \geq 0.05 \angstrom$, while $z=6\rightarrow5$ shows no change for these EWs. However, for $\mathrm{EW} < 0.05 \angstrom$ there is a pronounced decrease in the number density of $\oi$ absorption systems between each integer redshift from $z=7\rightarrow5$, with the strongest evolution occurring for $\mathrm{EW} < 0.08 \angstrom$ between redshifts 7 and 6. In this redshift interval, the total $\ell\left(X\right)$ decreases by 0.45, compared to 0.02 and 0.1 in $z=8\rightarrow7$ and $z=6\rightarrow5$, respectively. The change in EW-dependent $\ell\left(X\right)$ from $z=7\rightarrow6$ also becomes larger for smaller EW. For $\mathrm{EW} \geq 0.1 \angstrom$, $\Delta \ell\left(X\right) = 0.02$, but for $\mathrm{EW} \geq 0.03 \angstrom$ this difference increases to 0.13.

For comparison purposes, we plot observations from~\citet{becker19} divided into two of their $\Delta z=0.8$ redshift bins, $4.9 < z < 5.7$ and $5.7 < z < 6.5$. To approximate the effect of the uncertainty on each EW measurement, we add a random value selected from a Gaussian distribution with standard deviation equal to the reported uncertainty (see Table 3 of~\citet{becker19}). We take the 16th and 84th percentile values after repeating this process 1000 times and treat them as the lower and upper bounds of the observational distributions, giving rise to the vertical spread seen in Figure~\ref{fig:EW_OI}. While the observational curves are essentially identical for $\mathrm{EW} \geq 0.3 \angstrom$, there is a marked decrease in number density for the $\mathrm{EW} < 0.3 \angstrom$ range. The difference between the two curves also increases as $\mathrm{EW}$ becomes smaller, increasing from $\Delta \ell \left(X\right) = 0.16$ for $\mathrm{EW} \geq 0.1 \angstrom$ to $\Delta \ell \left(X\right) = 0.26$ for $\mathrm{EW} \geq 0.03 \angstrom$. Between these two redshift bins, the total $\ell\left(X\right)$ drops from 0.53 to 0.25, more than a factor of 2.

Examining the $z=6$ simulated distribution and the $5.7 < z < 6.5$ bin from~\citet{becker19} (labeled $z\approx6$ in the figure), the simulations \textit{under}predict $\ell \left(X\right)$ for $\mathrm{EW} \geq 0.1 \angstrom$ by a factor of $\approx 2$, but \textit{over}predict the abundance of $\mathrm{EW} \geq 0.03 \angstrom$ absorbers by a factor of 1.4. Despite these discrepancies, the simulations nearly perfectly reproduce the number of absorbers with $\mathrm{EW} \geq 0.05 \angstrom$. Considering the $4.9 < z < 5.7$ bin in~\citet{becker19} ($z\approx5$), the simulated curve is a factor of 1.5 too low for the $\mathrm{EW} \geq 0.1 \angstrom$ systems, but overpredicts the weaker systems with $\mathrm{EW} \geq 0.03 \angstrom$ by a factor of 2.3. By $z=5$, the incidence of $\mathrm{EW} \geq 0.05 \angstrom$ systems is also overpredicted. Although our method of accounting for uncertainty in the observations contributes some spread in the EW distribution, it is quite minor and does not bring the observations into better alignment with the simulated distributions.

To be certain the disparity is not a result of our long sightlines, we account for cosmic variance at $z=6$ by dividing our simulated sightline into coherent segments that each span an absorption pathlength $dX = 63.3$, comparable to the size of the~\citet{becker19} survey at $z=6$ (calculated using our cosmology), and plot the range of EW distributions in the blue hatched region in Figure~\ref{fig:EW_OI}. Even when accounting for this scatter, the simulated distribution remains somewhat too steep in comparison to their $z\approx6$ sample.

\begin{figure}
\centering
\includegraphics[width=0.5\textwidth]{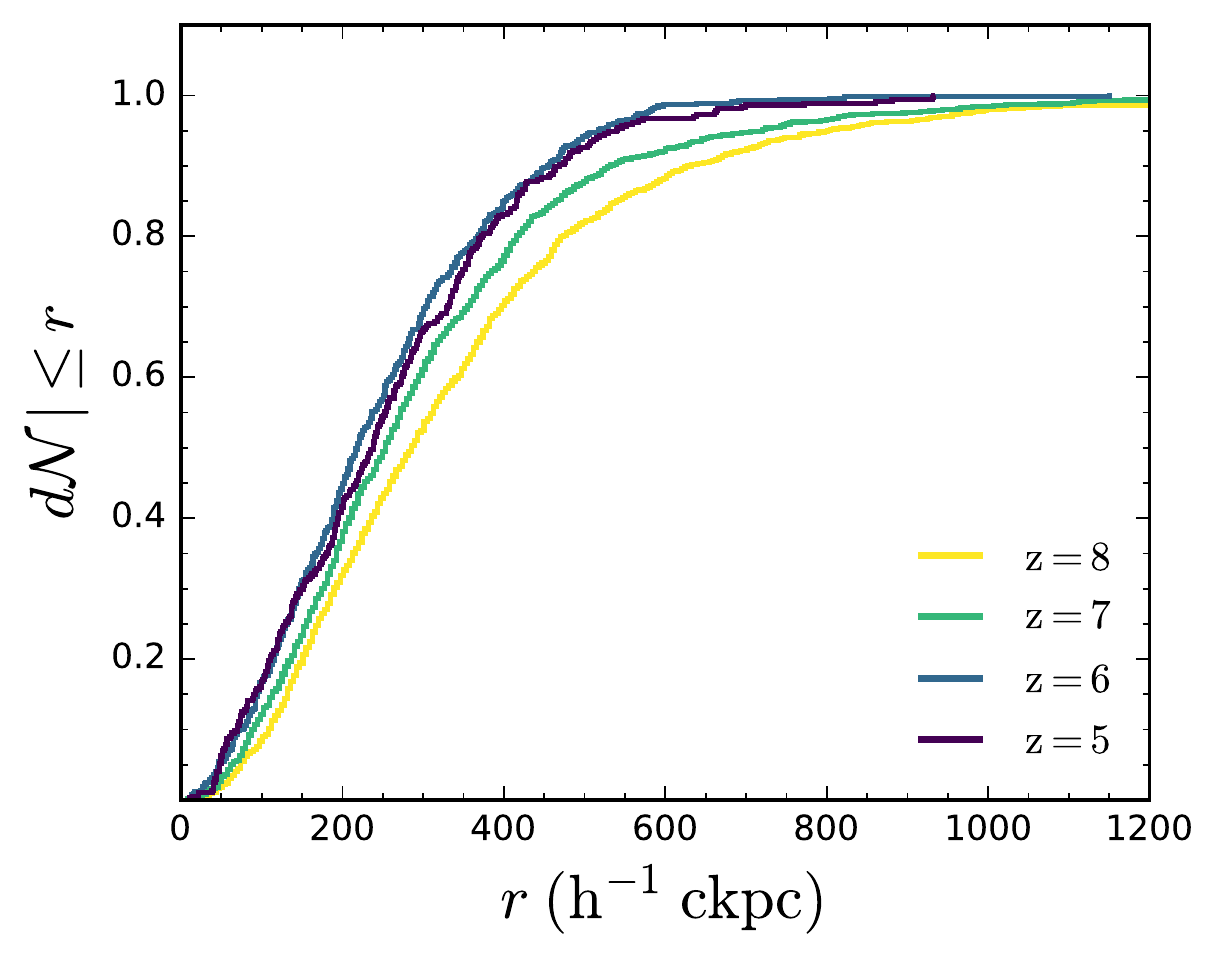}
\caption{Cumulative distribution of comoving distances of absorbers for different redshifts. Generally as the redshift decreases, the distribution peaks at a closer distance to the nearest halo. The largest deviation between redshifts occurs for distances less than roughly 360 $\hckpc$ (about $51 \; \hpkpc$ at $z=6$).\label{fig:changing_distance_withz}}
\end{figure}

\begin{figure}
\centering
\includegraphics[width=0.5\textwidth]{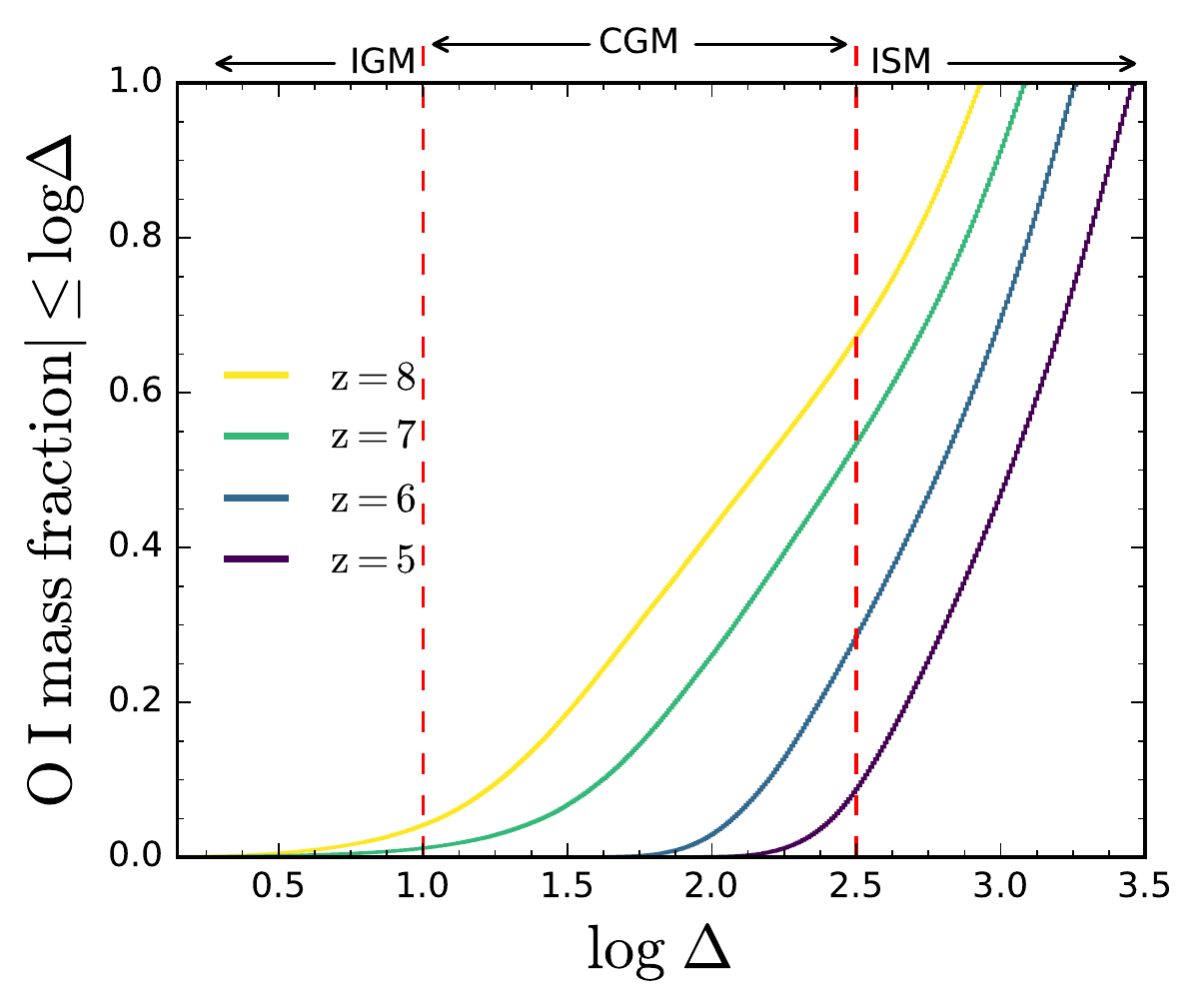}
\caption{The mass fraction of $\oi$ as a function of the overdensity for all enriched particles in the simulation, plotted for four different redshifts. The red dashed lines indicate the approximate density boundaries between the IGM, CGM, and ISM.}\label{fig:OI_mass_fraction}
\end{figure}

\subsection{O I absorber proximity to halos}
Noting the shrinking of the covering fractions in Section~\ref{ssec:spatial_distribution}, it raises the question of whether the typical separation between $\oi$ absorbers and their galactic hosts may change during this epoch as well. ``Host'' here refers to the halo physically closest to an $\oi$ absorber (as opposed to the halo with the smallest impact parameter from the sightline). Physically, as reionization progresses and the photoionization rate increases, the density threshold for gas to be self-shielded increases. The remaining neutral pockets will then be sheltered in these higher density regions and found preferentially closer to halos~\citep{miralda00}. One should then expect that the $\oi$ absorbers will become more closely associated with halos a s reionization progresses.

We plot the cumulative distribution of distances of $\oi$ absorbers versus their distance from the nearest halo, at $z=8\rightarrow5$ (Figure~\ref{fig:changing_distance_withz}). For $z=8$, the distribution reaches 50 percent of all absorbers by a distance of 290 $\hckpc$ and 100 percent by 1560 $\hckpc$ (although the graph is truncated at $r=1200 \hckpc$). For $z=7$ and 6 the distribution becomes progressively steeper, reaching 50 and 100 percent at smaller distances. For $z=5$ there is an exception, however, as the distribution ``rebounds'' a bit and the typical distance of $\oi$ absorbers increases very slightly. The typical distances at $z=5$ remain lower than those of $z=8$ and 7, however, so the long-term trend holds. By $z=5$, 50 percent of $\oi$ absorbers are within $235\;\hckpc$ and 100 percent are within $930\; \hckpc$. At $z=8$ then, $\oi$ systems are probing comoving distances from halos that are 1.6 times larger than at $z=5$. The differences in physical distances are not as stark, however, with the decrease amounting to only a factor of 1.1.

To explicitly see whether the $\oi$ is arising from the CGM or IGM, we examine the cumulative $\oi$ mass distribution as a function of overdensity within enriched particles in the simulation (Figure~\ref{fig:OI_mass_fraction}). The overdensity is calculated as the density of the gas particle over the mean density at the stated redshift, calculated as $\overline{\rho}=\Omega_b \rho_\mathrm{crit} \left(1+z\right)^3$.

At $z=8$, only a few percent of the total $\oi$ mass is located in the IGM, with roughly 60 percent of the mass located in overdensities considered characteristic of the CGM. By $z=7$, the mass fraction of $\oi$ in the IGM has dropped to approximately zero. As reionization progresses, the cumulative distribution retains roughly the same shape, perhaps becoming a little steeper, but shifts towards higher overdensities. At $z=6$ only 30 percent of the $\oi$ mass is contained in the CGM and by $z=5$ this has dropped further to roughly 10 percent, the remaining 90 percent occupying overdensities characteristic of the ISM.

\begin{figure}
\centering
\includegraphics[width=0.5\textwidth]{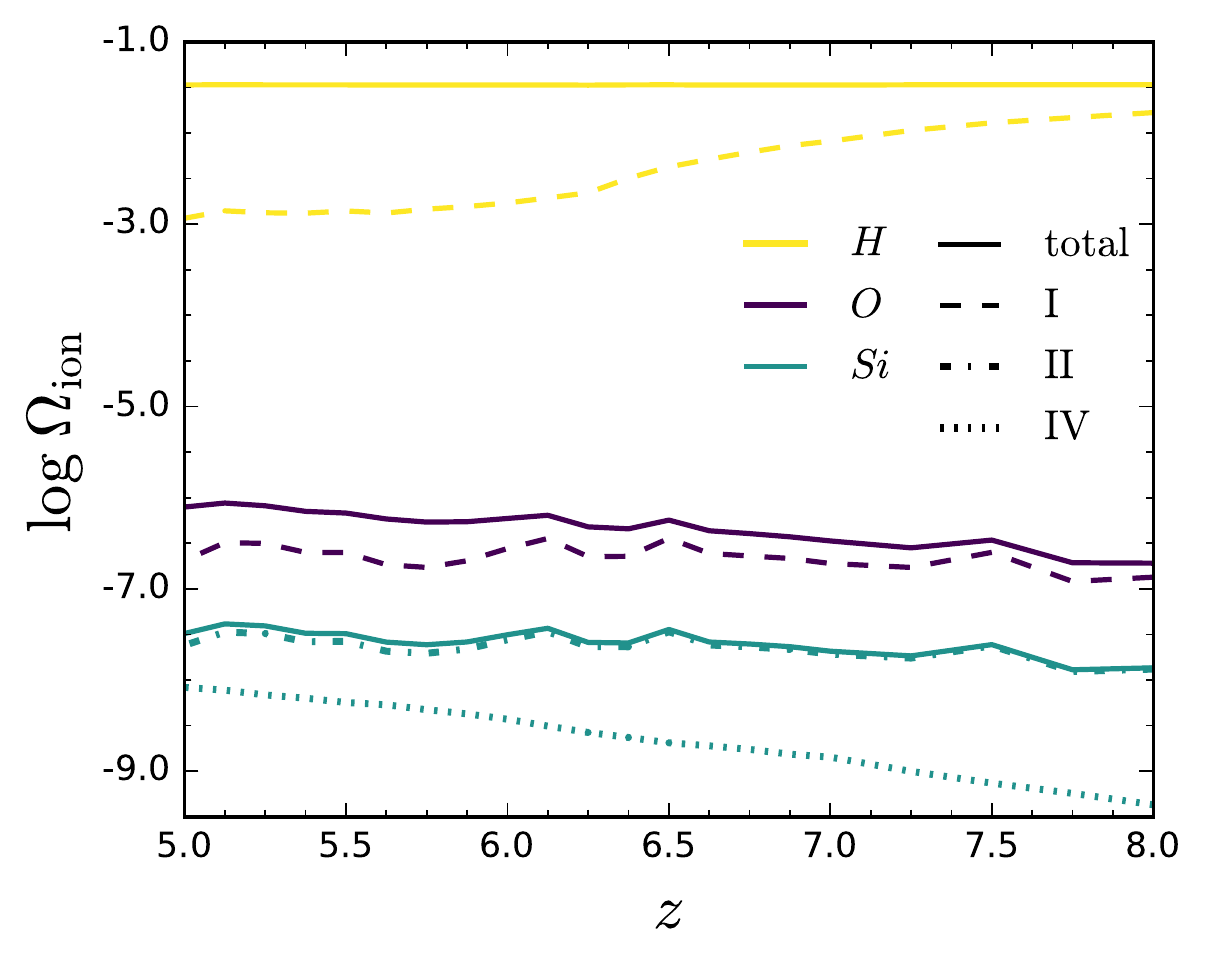}
\caption{Comoving mass densities of several species calculated along the sightline. The solid lines indicate the total comoving mass density for an element, while the other styles indicate that for an individual ionization state. Hydrogen is indicated in yellow, oxygen in purple, and silicon in blue. The neutral states of both hydrogen and oxygen gradually diverge from the total comoving mass density of each element as redshift decreases. The total silicon abundance, while much lower than that of oxygen, increases at a similar rate.\label{fig:comoving_mass_density}}
\end{figure}

\subsection{Comoving mass densities}\label{sssec:omega_comparison}
Considering the similar ionization energies of $\oi$ and $\hi$, it is reasonable to think the neutral fractions of these elements may evolve similarly through the EoR. In this section, we examine how the comoving mass densities of several species vary with respect to one another. Overall, we expect the fraction of neutral oxygen to decrease as reionization progresses due to the similar ionization energy of $\oi$ to that of neutral hydrogen. For all species in this section, the comoving mass density $\Omega$ is calculated as
\begin{equation}\label{eq:comoving_mass_density_calc}
\Omega_\mathrm{ion} = \frac{H_0 \; m_\mathrm{ion}}{c \; \rho_{\mathrm{crit}}} \frac{\sum_{i=1}^{n_{\mathrm{pix}}} N_\mathrm{ion,i}}{dX\left(z\right)}
\end{equation}
where $m_\mathrm{ion}$ is the mass of the oxygen ion and $\rho_{\mathrm{crit}}=9.2\times10^{-30}\;\mathrm{g\cdot cm^{-3}}$ is the critical density. $N_{\mathrm{ion},i}$ is the column density of the ion in that ``pixel'' of the sightline, while $n_\mathrm{pix}$ is $2 \times 10^6$, the number of pixels along the sightline. The oxygen abundance, $\Omega_{O}$, is calculated by summing the column densities in each pixel of the first four ionization states of oxygen (O I $\text{--}$ O IV). Only the first four states are included in the calculation of the total oxygen because more highly ionized states, such as $\ov$ and $\ovi$, are preferentially located around hot halos or in regions with an exceptionally hard UVB, both of which are rare for $z>6$.

We plot the comoving mass densities of several species in oxygen and hydrogen (Figure~\ref{fig:comoving_mass_density}). We also include silicon to compare its evolution in redshift to that of oxygen (see subsection~\ref{ssec:xHI_Omega_ratios} for discussion of the motivation). While the overall hydrogen holds steady, $\hi$ shows strong evolution from $z=8\rightarrow5$, decreasing over this range. From $z=8\rightarrow5$, $\Omega_{\oxygen}$ gradually increases by about 0.5 dex. The comoving mass density of $\oi$ follows the same general trend, increasing in the long-term while approximately paralleling the overall oxygen abundance, down to roughly $z=6.5$. Below this redshift, $\Omega_{\oi}$ diverges more quickly to lower values as the oxygen is pushed into more highly ionized states and the overall $\oi$ abundance starts to decrease. $\Omega_{\si}$ increases in a similar manner to $\Omega_{\oxygen}$, although its increase is less than 0.5 dex from $z=8\rightarrow5$, slightly weaker than that of oxygen.

Although the specific trends of neutral and total hydrogen and oxygen differ from one another, there appears to be a decrease in the ratio of $\Omega_{\oi}$ with respect to $\Omega_\oxygen$ as there is a concurrent decrease in $\Omega_{\hi}$ with respect to $\Omega_\hydrogen$. The similarity of these two quantities' evolutionary trends suggests there may be a way to relate the two, a venture which we pursue in the following two subsections.

\subsection{O I incidence rate and volume-weighted neutral hydrogen fraction}\label{sssec:lX_xHIv}

\begin{figure}
\centering
\includegraphics[width=0.5\textwidth]{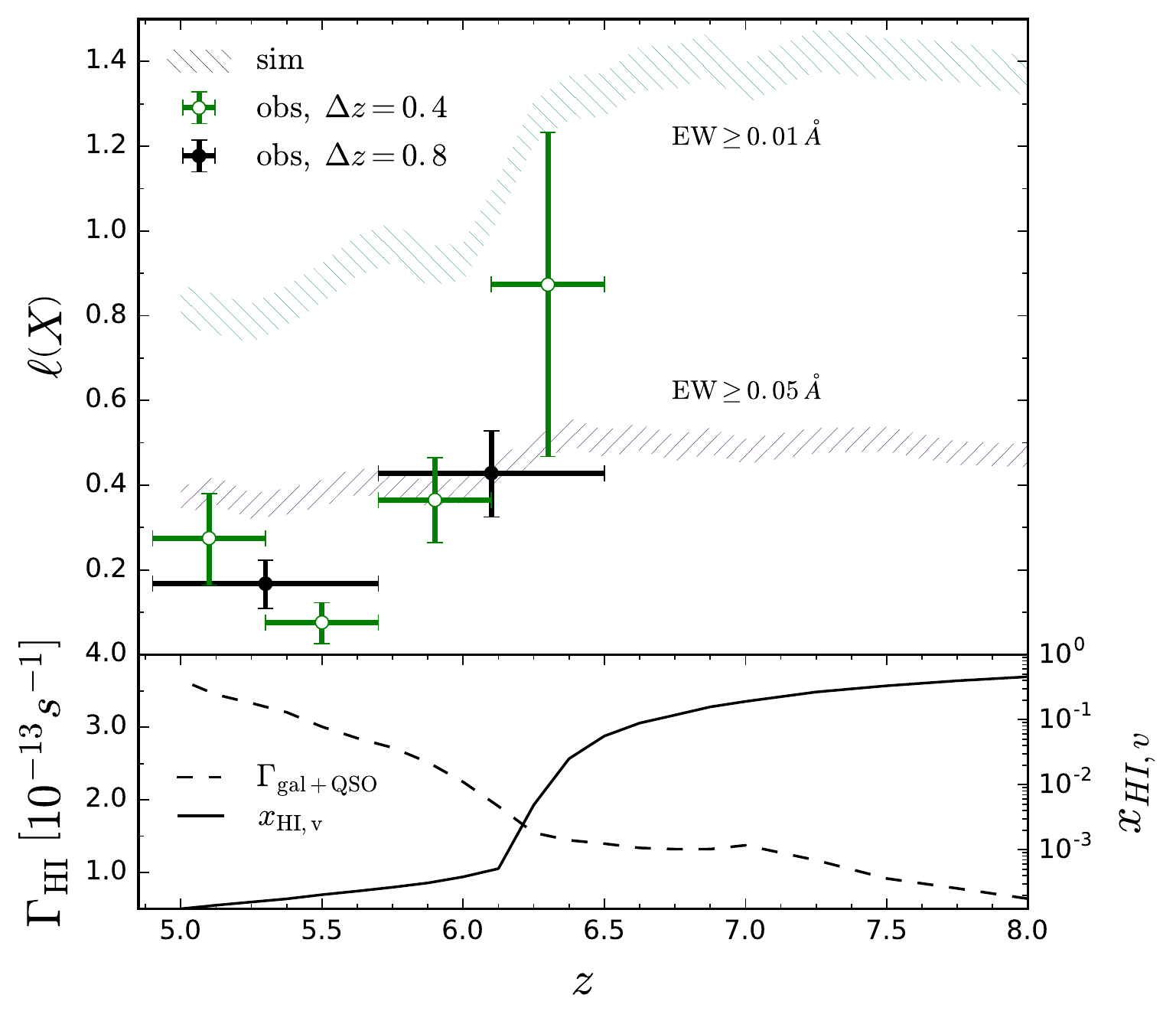}
\caption{\textit{Top:} The evolution in the incidence rate of $\oi$ absorbers. The uncertainty in $ \ell \left(X\right) $ is given as $\sqrt{n}$ where $n$ is the number of absorbers detected in the simulation at the given redshift. Two cutoffs in equivalent width have been applied: one at $\mathrm{EW} = 0.01 \angstrom$ (blue hatching) and another at $\mathrm{EW} = 0.05 \angstrom$ (purple hatching). The black (green) points show $\ell\left(X\right)$ from~\citet{becker19} in redshift bins of $\Delta z = 0.8$ ($\Delta z = 0.4$), with the vertical error bars indicating the 68 percent confidence level. The data include an EW-based completeness correction. \textit{Bottom:} Evolution of the globally calculated neutral hydrogen fraction on a vertical log scale (solid black) overplotted with the volume-averaged $\hi$ photoionization rate, with contributions from both galaxies and quasars (dashed black). \label{fig:incidence_rate_xHIv}}
\end{figure}

\begin{figure*}
\centering
\includegraphics[width=1.0\textwidth]{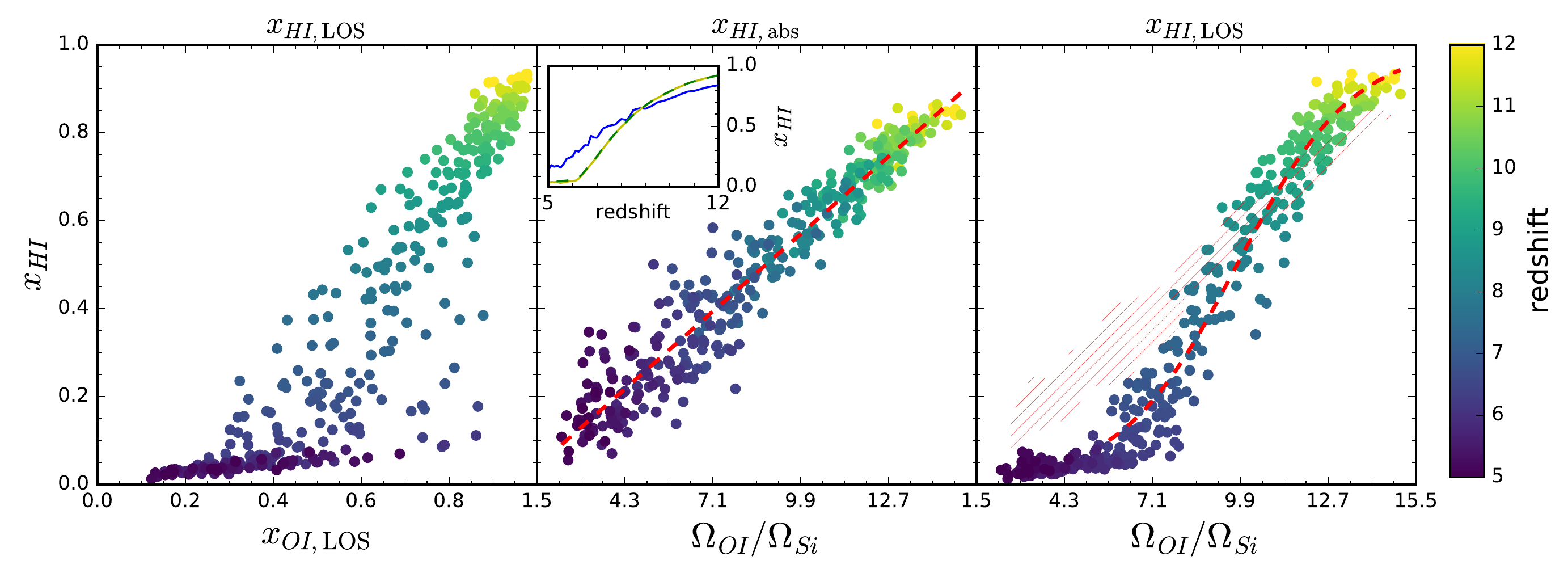} \caption{\textit{Left panel:} The neutral fractions of $\hi$ and $\oi$, each calculated as ratios between the comoving mass density of the neutral state to that of the element, plotted for all snapshots from $z=12\rightarrow5$. The values are calculated using the column density of $\hi$ and $\hii$ and the first four states of oxygen. All particles encountering the drawn sightline are included in the calculation, regardless of whether they are associated with an identified absorber. \textit{Middle panel:} The calculated $\hi$ neutral fraction when contributions are restricted to particles belonging to absorption systems, plotted against the ratio of $\Omega_{\oi}$ to $\Omega_{\si}$, our suggested proxy for $x_{\oi}$. The inset figure shows the median $x_{\hi, \mathrm{abs}}$ predicted by the absorbers on each sightline segment (solid blue) and $x_{\hi, \mathrm{LOS}}$ (solid yellow) with the true global mass-weighted value (dashed green) against redshift. \textit{Right panel:} The neutral hydrogen fraction calculated as in the left panel, plotted against $\Omega_{\oi}$ to $\Omega_{\si}$. The red dashed line indicates the fit line described in Equation~\ref{eq:xHIm_OmegaOISi_fit}. The red hatching shows the $1\sigma$ distribution about the median of points in the middle panel for comparison.\label{fig:xHI-proxies}}
\end{figure*}

We begin by comparing the evolution in the simulated incidence rate of absorbers, $ \ell \left(X\right) $, to the volume-averaged neutral hydrogen fraction, $x_{\hi,v}$  (Figure~\ref{fig:incidence_rate_xHIv}). To calculate the incidence rate, $\oi$ absorbers for each redshift are tallied from the sightline catalog and the number is divided by the pathlength, while the global $x_{\hi,v}$ is taken directly from the simulation output. We analyze two curves: one with a minimum EW cutoff of $\mathrm{EW} = 0.05 \angstrom$ to compare to the observations of~\citet{becker19} and another with cutoff $\mathrm{EW} = 0.01 \angstrom$ to illustrate the incidence rate's dependence on EW.

Considering an EW cutoff of $\mathrm{EW} \geq 0.05$, from $z=8\rightarrow6.3$, the incidence increases slightly, from $\ell \left(X\right)=0.47$ to 0.48. Below $z=6.3$, it drops to 0.38 by the time the simulation reaches $z=5.9$, a factor of 1.26 change. It continues to gradually decrease to 0.36 at $z=5.0$. Overall, by $z=5$ the number density of $\mathrm{EW} \geq 0.05 \angstrom$ $\oi$ systems has decreased by 23 percent, with the majority of this evolution occurring between $z=6.3$ and 5.9. The incidence rate for $\mathrm{EW} \geq 0.01 \angstrom$ systems shows a similar general trend. At $z=8.0$, it has a value of $\ell \left(X\right)=1.36$ that increases very slightly up to $\ell\left(X\right) = 1.39$ at $z=6.8$ before undergoing a slow decrease that becomes more rapid at $z=6.3$ and ends at $z=5.9$ after decreasing by $\Delta \ell\left(X\right)=0.33$. Below this redshift, there continues to be a slow decrease in incidence rate before it settles at $\ell \left(X\right) =0.82$ by $z=5$. In total, from $z=8\rightarrow5$ the incidence rate for this smaller EW cutoff shrinks by 40 percent, a noticeably larger change than is seen in our $\mathrm{EW} \geq 0.05 \angstrom$ sample.

\citet{becker19} evaluates their $\mathrm{EW} \geq 0.05\angstrom$ data in two bin sizes in redshift: $\Delta z=0.8$ and $\Delta z=0.4$. For their $\Delta z=0.8$ bins they calculate the most likely value of $\ell \left(X\right)$ at the 68 percent confidence level, finding $\ell\left(X\right) = 0.421^{+0.098}_{-0.101}$ for $5.7 < z < 6.5$ and $0.165^{+0.055}_{-0.058}$ for $4.9 < z < 5.7$. This gives a change in $\ell \left(X\right)$ of a factor of $2.5^{+1.6}_{-0.8}$ at 68 percent confidence occurring across $z=5.7$. Using the smaller $\Delta z = 0.4$ bins, the change may be even greater, although this is dominated by the $ 6.1 < z < 6.5$ redshift bin, which has very large associated uncertainties due to the low $\oi$ detection count (only 5 systems). In either case, the observed evolution near $z\approx 6$ occurs at slightly later redshifts compared to the simulation, and is more extreme than that of our equivalent EW cutoff.

Throughout the $z=8\rightarrow6.3$ regime, the volume-weighted neutral hydrogen fraction, $x_{\hi,v}$, decreases slowly before falling precipitously below $z=6.5$ (bottom panel of Figure~\ref{fig:incidence_rate_xHIv}). It falls below 1 percent at $z=6.25$ and by $z=6$ is effectively zero and reionization is complete in the simulation. While the specifics of the trends of $x_{\hi,v}$ and $\ell \left(X\right)$ are different, they both demonstrate a change in evolutionary behavior near $z\approx6.3$: At roughly the same time, hydrogen becomes nearly completely ionized and the $\oi$ number density starkly decreases. This could indicate that trends in $\oi$ incidence with redshift may be used as a tracer of certain stages of $\hi$ reionization.

\subsection{Neutral hydrogen fraction and comoving density ratios in oxygen and silicon}\label{ssec:xHI_Omega_ratios}
To further clarify this potential relationship, we directly compare the local neutral fraction of hydrogen with that of oxygen\footnote{``Local'' here refers to individual sightline segments, and we indicate quantities calculated in this way using subscript ``LOS'' (for line-of-sight) to distinguish between this metric and a volume-weighted or mass-weighted global quantity, or one calculated exclusively from particles contributing to absorption systems (subscripted ``abs'' for absorber).}, calculating each from segments of the sightline in all simulation snapshots ranging from $z=12$ to $z=5$, where the unweighted neutral fraction is calculated as a ratio of comoving mass densities (left panel of Figure~\ref{fig:xHI-proxies}). The segments are created by dividing the entire $4 \times 10^6$ km/s sightline into ten equal pieces in velocity space ($dX \approx 44$ at $z=6$ using our cosmology) and considering each as an independent sightline with its own collection of absorption systems.

For $x_{\oi, \mathrm{LOS}} < 0.3$, there is a tight relationship with a very shallow positive slope. Pushing to higher $x_{\oi, \mathrm{LOS}}$, there appears to be a general trend with a steep upward slope extending from roughly $x_{\oi, \mathrm{LOS}} \approx 0.4\rightarrow 0.9$. However, the scatter in the relationship is considerably broader at higher $x_{\oi, \mathrm{LOS}}$ and there are many sightline segments that show an atypically low $x_{\hi, \mathrm{LOS}}$ for a given $x_{\oi, \mathrm{LOS}}$. The range in possible $x_{\hi, \mathrm{LOS}}$ values for higher $x_{\oi, \mathrm{LOS}}$ spans nearly 0.8.

It is important to note that the usage of $x_{\oi}$ as a proxy for $x_{\hi}$ in an observational context requires a method of measuring the total oxygen column density, which would necessitate measurements of additional metal absorbers in tandem with ionization modeling using codes such as {\sc cloudy}~\citep{ferland98}. To remove the necessity of further modeling, we replace the total oxygen density $\Omega_\mathrm{O}$ with $\Omega_\mathrm{Si} \left(\equiv \Omega_{\siii}+\Omega_{\siiv}\right)$.\footnote{We assume the total comoving density in silicon can be approximated as the sum of singly-ionized and triply-ionized silicon, as these two species dominate the abundance in the simulation.} Silicon is selected over other metals for its tendency to evolve similarly to oxygen, though oxygen is roughly an order of magnitude more abundant. Additionally, several transitions in silicon are readily observable in quasar absorption spectra.

We examine the relationship between $x_{\hi, \mathrm{abs}}$ and $\Omega_{\oi}/\Omega_{\si}$, hereafter referred to as $\Omega_{\oi \si}$ (middle panel of Figure~\ref{fig:xHI-proxies}). Deviating from the original treatment, which included all gas contributions to the column density in the calculation of $\Omega$ for each transition, we now restrict the total column density for both $\Omega_{\oi\si}$ and $x_{\hi}$ in each segment to include only gas within the detected metal line absorption systems, in effect altering Equation~\ref{eq:comoving_mass_density_calc} to become
\begin{equation}\label{eq:comoving_mass_density_absorbers}
\Omega_\mathrm{ion} = \frac{H_0 \; m_\mathrm{ion}}{c \; \rho_{\mathrm{crit}}} \frac{\sum_{i=1}^{n_{\mathrm{abs}}} N_\mathrm{ion,i}}{dX\left(z\right)}
\end{equation}
Restricting the contributions to detected metal absorption systems better mimics the process used by observers to make measurements of the comoving density. In addition to this change, we also introduce lower limit cutoffs in the column densities of the absorbers contributing to the $\Omega$ calculation: $\log N_{\oi}=13.49$, $\log N_{\siii} = 12.11$, and $\log N_{\siiv} = 12.43$. The $\oi$ cutoff is the minimum observed column in~\citet{becker11}, while the cutoffs in $\siii$ and $\siiv$ are calculated as the column density in these transitions with equivalent optical depth to the $\oi$ cutoff.  We plot the relationship between these quantities in the middle panel of Figure~\ref{fig:xHI-proxies}.

Applying these restrictions to the calculation of $\Omega_{\oi\si}$ and $x_{\hi, \mathrm{abs}}$ reveals a strong linear relationship between the two quantities, described well by a simple equation:
\begin{equation}\label{eq:linear_proxy_fit}
x_{\hi,\mathrm{abs}} = 0.06 \; \Omega_{\oi\si} - 0.05
\end{equation}
As evident in the figure, the scatter in the relationship is quite small for $z>8$ and then increases for lower redshifts. However, the fit line predicts $x_{\hi, \mathrm{abs}}$ to within 10 percent of the true value in absorbers for 95 percent of the sightline segments, suggesting a strong correlation between the two quantities. This relationship persists even in the face of increased scatter as patchy reionization causes prominent spatial fluctuations in the local $\hi$ neutral fraction for lower redshifts.

Next we investigate the extent to which this relationship is retained when considering the $x_{\hi}$ fraction calculated along the entirety of each sightline segment. This explicit examination is necessary because gas particles associated with absorption systems are predisposed to occupy unusually neutral gas as redshift decreases, meaning that estimates of the neutral hydrogen fraction made using Equation~\ref{eq:linear_proxy_fit} would suggest a peculiarly high neutral fraction. For example, at $z=7.0$ the global mass-weighted $x_{\hi}$ is 0.25, while the mean $\Omega_{\oi\si}=7.37$. The fit function would then imply a much larger value of $x_{\hi} = 0.39$. This deviation only becomes more stark for lower redshifts.

We compare $x_{\hi, \mathrm{LOS}}$ to $\Omega_{\oi\si}$ in the right panel of Figure~\ref{fig:xHI-proxies}, where the scatter is drastically reduced compared to the relationship between $x_{\hi, \mathrm{LOS}}$ and $x_{\oi, \mathrm{LOS}}$. For $x_{\hi, \mathrm{LOS}} > 0.1$, there is a positive correlation with $\Omega_{\oi \si}$, whereas for $x_{\hi, \mathrm{LOS}} < 0.1$, the slope relating the two values becomes much shallower. The scatter in $x_{\hi, \mathrm{LOS}}$ for a given $\Omega_{\oi \si}$ also decreases for these smaller values of $x_{\hi, \mathrm{LOS}}$. Compared to the previous test, the relationship is decidedly less linear, resembling instead a sigmoid function. Performing optimization on the relationship for $x_{\hi}>0.1$ while assuming a sigmoid function results in the following fit line
\begin{equation}\label{eq:xHIm_OmegaOISi_fit}
x_{\hi, m} = \frac{1}{1+\exp \left(-0.51 \left(\Omega_{\oi\si} - 9.73\right) \right)} \: \mbox{for } \Omega_{\oi\si} > 5.42
\end{equation}
There is a maximum scatter of roughly $\pm0.2$ about the fit line, although more than 90 percent of the sightline segment $x_{\hi, \mathrm{LOS}}$ values fall within 10 percent of the predicted value. 

The relatively small scatter in the relationship here is encouraging, but since both $x_{\hi, \mathrm{LOS}}$ and $\Omega_{\oi \si}$ are calculated as a piece of a sightline, which only directly probes gas along the line of sight, it is worth considering whether $x_{\hi, \mathrm{LOS}}$ values derived using Equation~\ref{eq:xHIm_OmegaOISi_fit} are representative of a global measure of the hydrogen neutral fraction, for example the mass-weighted neutral fraction, $x_{\hi,m}$. We include an inset plot in the middle panel of Figure~\ref{fig:xHI-proxies} showing the median values of $x_{\hi,\mathrm{abs}}$ and $x_{\hi,\mathrm{LOS}}$ from $z=12\rightarrow5$ overplotted with those of the global $x_{\hi,\mathrm{m}}$. We find that the median $x_{\hi,\mathrm{LOS}}$ of each sightline segment calculated lies directly over the line of global $x_{\hi,\mathrm{m}}$ for our entire studied range of redshifts, suggesting that the true mass-weighted $\hi$ fraction could be estimated using a relationship similar to Equation~\ref{eq:xHIm_OmegaOISi_fit}. However, $x_{\hi, \mathrm{abs}}$ underpredicts the global mass-weighted neutral fraction for $z>8$ and overpredicts for $z<8$. It diverges more strongly for $z<8$ but always remains within 20 percent of the true value. 

\section{Discussion}\label{sec:discussion}
We begin our discussion by addressing the three questions that were raised in Section~\ref{sec:introduction}, making particular comparison to two theoretical works. \citet{finlator13} used cosmological hydrodynamic simulations to learn about the $\hi$ hosts of $\oi$ and to study the enrichment of the CGM and IGM. \citet{keating14} similarly used cosmological hydrodynamic simulations (originally described in~\citet{bolton13}) to study the association between $\oi$, gas metallicity, and $\hi$ column densities during reionization.

\subsection{How do the neutral oxygen halos of galaxies evolve with time?}
From $z=7\rightarrow6$, there is a decrease in the incidence rate for systems with $\mathrm{EW} < 0.08 \angstrom$, indicating that it is the weak absorption systems that show the most change during reionization. Systems with $\mathrm{EW} > 0.08 \angstrom$ show a decrease in incidence rate from $z=8\rightarrow7$, but it is relatively small and does not proceed to lower redshifts. The affected low EW systems are also arising predominately from lower relative overdensities. This picture is consistent with previous studies demonstrating an inverse correlation between absorber column density and impact parameter from a host halo (see e.g.~\citet{garcia17}). Although not shown here, we confirmed that this trend is reproduced in our simulations by directly comparing typical column density with distance from the nearest host halo for the simulated $\oi$ systems. Our $\oi$ systems fall in the linear portion of the curve of growth, so there is a positive correlation between the equivalent width and the column density.

This migration of $\oi$ systems to denser gas nearer to halos implies also that the metallicity of the absorber-hosting gas increases. In conceptual agreement with our results are those of \citet{keating14}, who made comparison of the metallicities of their $\oi$ population ($\mathrm{EW} > 0.01 \angstrom$) to those of Lyman limit systems and damped Ly$\alpha$ absorption (DLA) systems at lower redshift~\citep{simcoe12,rafelski12}. The overall trend presented in the DLA observations was a decreasing metallicity with increasing redshift, with the simulated metallicity of the~\citet{keating14} $\oi$ systems falling neatly within the observed trend. It seems to suggest that $\oi$ absorbers arise within more metal-poor gas as redshift increases, which make sense intuitively considering the ongoing process of enrichment. Thus, a scenario is presented in which the $\oi$ distribution around halos shrinks as the redshift decreases, changing most dramatically below $z=7$.

\subsection{Is neutral oxygen found predominantly in the IGM or the CGM?}
In order to address this question, we have separated gas into ISM, CGM, and IGM by overdensity: ISM is any gas with $\log \Delta > 2.5$; CGM denotes gas with $10 < \log \Delta \leq 10^{2.5}$; the IGM has $\log \Delta \leq 10$~\citep{pallottini14}. Our analysis shows that the majority by mass, $\approx$ 65 percent, of the $\oi$ gas appears in overdensities characteristic of the CGM for $z=8$ and $z=7$. The full distribution of overdensities in the simulation at these redshifts ranges from $-0.5 \leq \log \Delta \leq 2.9$, extending from the IGM into the ISM. By $z=6$, the majority of $\oi$ mass has shifted to even higher densities typically associated with ISM gas, with less than 30 percent of the total $\oi$ mass falling in CGM overdensities, $\log \Delta \leq 2.5$. By $z=5$, this has dropped even further to a mere 10 percent.

\citet{keating14} found that $\oi$ systems in their simulations occupied overdensities of $\log \Delta > 1.69$, especially in the range $1.69 < \log \Delta < 1.90$. This essentially places \textit{all} of their simulated absorbers in the CGM. Further, they found that their simulated $\oi$ systems arise in neutral hydrogen columns ranging from $17.9 < \log N_{\hi} < 19.9$, falling into the LLS and sub-Damped Ly$\alpha$ system (sub-DLA) column density regime. Some simulations indicate that sub-DLAs are produced by both galaxy disk gas (analogous to the ISM) and galaxy halo gas (CGM), at least for $0.4 < z < 3$~\citep{rhodin19}. Based on these implications, the~\citet{keating14} $\oi$ systems may be forming in different galactic structures whose characteristic overdensities are analogous to ISM, CGM, and IGM gas.

\begin{figure}
\centering
\includegraphics[width=0.5\textwidth]{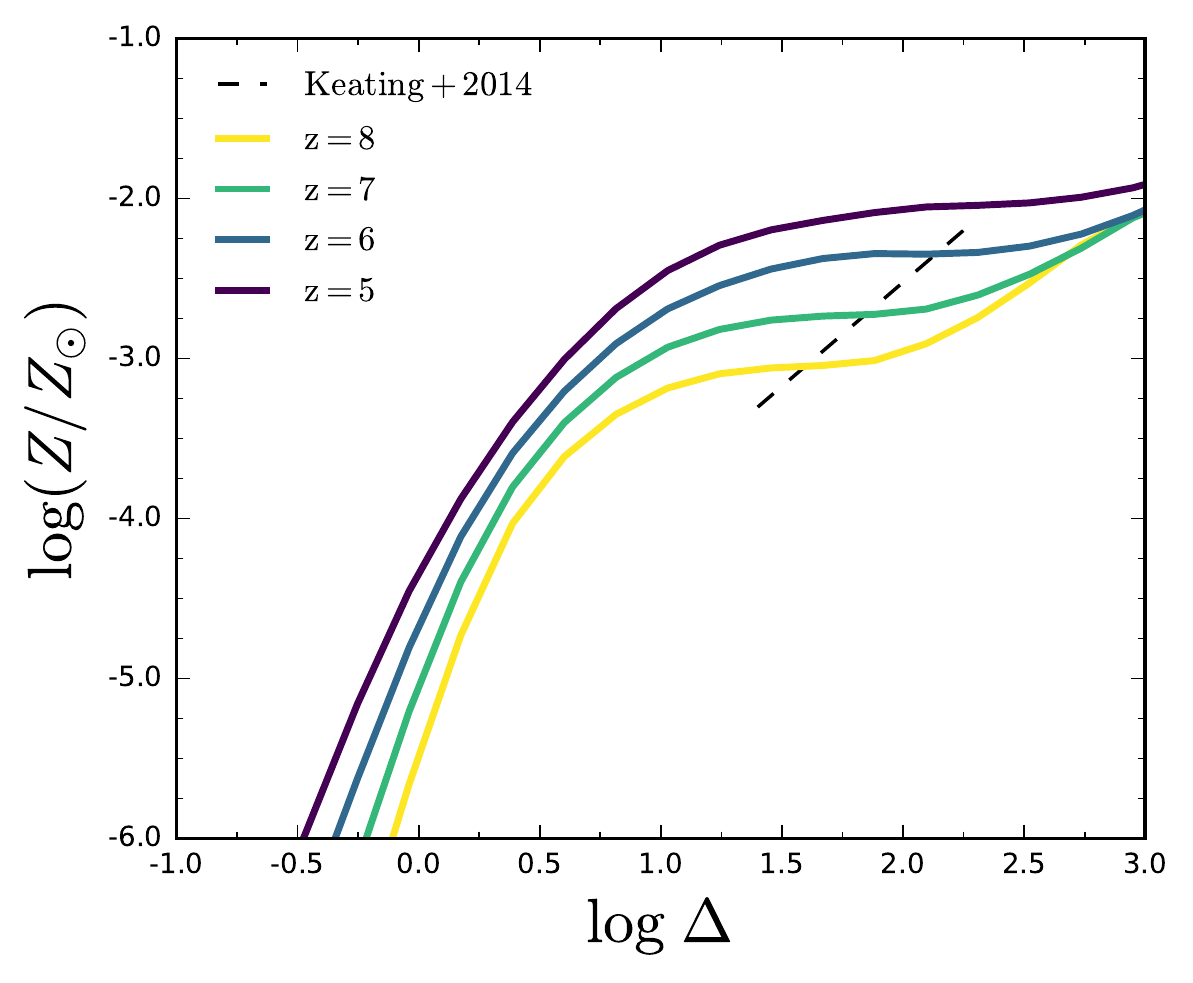}
\caption{The mean metallicity of all gas particles within the simulation for several redshifts, compared to that of~\citet{keating14} at $z=6$. The metallicity dependence predicted by our simulation agrees in normalization with their prescription although the slope is slightly shallower, leading to increased contribution to $\oi$ absorbers from less overdense gas.}\label{fig:meanlogZ_logDelta}
\end{figure}

Certain aspects of our findings differ from their results. Our $\oi$ is found in a wider range of overdensities, from $\log \Delta = -0.5$ (although these are quite rare) all the way up to $\log \Delta = 3.5$ with $\oi$ absorbers arising in progressively more overdense gas at later times. In terms of gas phase, this ranges from the comparatively dense ISM well out into the IGM. Apparently, the simulations of~\citet{finlator18} are more inclined to create neutral oxygen at higher overdensities when compared to the~\citet{keating14} framework. However, this difference is to be expected since the metals in ~\citet{finlator18} are modeled self-consistently whereas ~\citet{keating14} uses a power-law metallicity-density relation to apply metals in post-processing. To illustrate these differences, Figure~\ref{fig:meanlogZ_logDelta} shows the mean metallicity of all particles within our simulation for select redshifts from $5 \leq z\leq 8$ in comparison to the prescription used in~\citet{keating14}. The range of overdensities in~\citet{keating14} is significantly smaller by comparison, spanning only $\sim$one order of magnitude compared to our $\sim$four. Since we are modeling a substantial number of gas particles that are in the IGM but our $\oi$ absorbers are still restricted for the most part to the CGM, it seems reasonable to conclude that the diffuse IGM is not enriched enough to contribute significantly to the population of $\oi$ systems. This conclusion is bolstered both by the fact that the $\oi$ absorbers in~\citet{keating14} were hosted well inside the range of overdensities they considered, and since they were able to reproduce the observations of~\citet{becker11} quite well.

Our results fall more in line with those of \citet{finlator13}, which used a similar self-consistent metal enrichment scheme, where $\oi$ systems were found to be associated with $\hi$ column densities $N_{\oi} \sim 10^{19} - 10^{21}\; \mathrm{cm}^{-2}$. This column density range overlaps with the upper end of values found in~\citet{keating14}, but also extends higher into the DLA regime, which may preferentially be associated with the ISM within a galaxy disk~\citep{rhodin19}. Taking these theoretical works in conjunction with one another, the results appear to suggest that $\oi$ arises primarily in the CGM, particularly for redshifts above $z=6$. For $ 5\leq z \leq 6$, it may retreat into even higher overdensities associated with the ISM.

\subsection{Can the statistics of OI absorbers be used to trace the progress of hydrogen reionization?}
\subsubsection{The incidence rate}
We analyzed the redshift-evolution of the $\oi$ absorber incidence rate from $z=8\rightarrow5$ and determined that there is a drop in the number of $\oi$ absorption systems with $\mathrm{EW}\geq 0.01 \angstrom$ starting at $z\approx6.3$: $\ell \left( X\right)$ decreases from $1.25$ at $z=6.3$ to 0.92 by $z=5.9$. The same general trend is present when considering only $\oi$ systems with $\mathrm{EW} \geq 0.05 \angstrom$, although to a lesser degree. This evolution in $\oi$ is concurrent with an increase in the volume-averaged $\hi$ photoionization rate at $z=6.3$. The increase in the $\Gamma_{\hi}$ slope corresponds to the beginning of the ``bubble overlap'' phase of reionization, wherein the $\hii$ regions surrounding sources of ionizing radiation merge and the mean free path of hydrogen-ionizing photons in the IGM increases sharply. At the same redshift as this change in the ionization rate, the volume-weighted neutral fraction in hydrogen drops below 1 percent in the simulation, which is consistent with other theoretical studies of reionization~\citep{gnedin00}.

Given that $\hi$ and $\oi$ are sensitive to the same photons, and the simultaneous evolution in $\Gamma_{\hi}$ due to reduction in the IGM opacity, we conclude that these evolutionary phases are likely to occur at similar times. We further predict strong evolution of the $\oi$ absorber incidence when $\hii$ regions begin to overlap with one another and the neutral hydrogen fraction approaches 1 percent near the completion of reionization. Turning to the observational data, the $\oi$ systems of~\citet{becker11} were all detected at $z>5.75$, despite nearly half of their probed pathlength being below this redshift. Further, the expanded observations reported in~\citet{becker19} show a similar increase above $z=5.7$ (possibly an even stronger one above $z=6.1$ in their smaller redshift bins). Considering our results, if this apparent ``jump'' in the number of their $\oi$ absorbers is real, it may imply that the true volume-weighted neutral hydrogen fraction drops below 1 percent at $z\approx5.7$.

\citet{keating14} similarly evaluated the $\oi$ incidence rate, both cumulative and EW-dependent, in their simulations, comparing to samples of incidence rates of LLSs and DLAs at lower redshifts to investigate the origins of $\oi$-hosting structures~\citep{songaila10,fumagalli13,seyffert13}. Studying simulation outputs at $z=6$, 7, and 8 generated for a variety of plausible ionizing emissivities, they find an $\ell\left(X\right) \approx 0.8$ for $z=6$ for their fiducial ionization rate, $\log \Gamma = -12.8$. For higher redshifts their incidence rates increase, spanning a range of $\ell\left( X\right) \approx 1.7-2.1$ at $z=8$ for $\mathrm{EW} \geq 0.01 \angstrom$ with varying $\Gamma$. The number of absorbers increases by roughly a factor of $3-5$ when their cutoff equivalent width is decreased to $0.001 \angstrom$. They further find that the incidence rate of $\oi$ absorbers increases with redshift. These are all comparable to our findings. Also similar to our results, their predicted evolution in $\ell\left(X\right)$ appears to be more gradual than that observed in~\citet{becker11}.

\subsubsection{$x_{OI}$ and $\Omega_{OISi}$}
We first examined the relationship between the neutral fractions of hydrogen and oxygen in all particles falling along the simulated sightline segments. It is apparent that while there is roughly a linear trend between the two quantities, there is also a substantial amount of scatter. This is especially true for neutral hydrogen fractions between $0.1 \leq x_{\hi, \mathrm{LOS}} \leq 0.6$. While we believe that this is a realistic result, it doesn't accurately recreate an observing scenario, and thus doesn't provide much guidance or use to the observer.

In search of a more observationally useful metric, we explore the correlation between $x_{\hi}$ and a simple proxy for the neutral oxygen fraction, $\Omega_{\oi\si}$. Further, we initially restrict the contributions to $\Omega_{\oi\si}$ and $x_{\hi}$ to the particles in absorption systems identified in the simulation. A linear relationship is readily apparent when these restrictions are applied, showing that in principle $\Omega_{\oi\si}$ could serve as a highly constraining measurement on the evolution of neutral hydrogen during reionization. However, although redshift evolution in $x_{\hi, \mathrm{abs}}$ is evident, the implied neutral fraction is significantly higher than the global mass-weighted neutral hydrogen fraction. The discrepancy becomes more pronounced as the redshift decreases below $z=8$, roughly when $x_{\hi, v}$ reaches 50 percent. Below $z=8$, the IGM and CGM diverge, after which the neutral hydrogen fraction typical of metal absorption systems in the CGM becomes no longer representative of the simulation volume as a whole. 

Including all particles along the sightline segments in the neutral hydrogen fraction calculation shows a deviation from the previously tight linear correlation. For values $x_{\hi} \gtrapprox 0.1$, $x_{\hi,\mathrm{LOS}}$ appears to evolve more rapidly in time than $\Omega_{\oi \si}$. For $x_{\hi} \lessapprox 0.1$ the relationship is more linear, the slope becoming shallower and the scatter is decreased. The $x_{\hi} \gtrapprox 0.1$ regime is well-fit by Equation~\ref{eq:xHIm_OmegaOISi_fit}, with a relatively low scatter, but variables apart from the evolving ionizing background appear to be playing a part in the relative abundances of $x_{\hi, \mathrm{LOS}}$ and $\Omega_{\oi\si}$, causing a curvature in the relationship.

It may appear counterintuitive that $x_{\hi, \mathrm{LOS}}$ displays much more scatter with $x_{\oi, \mathrm{LOS}}$ than with $\Omega_{\oi \si}$, since the latter has less of a physical motivation to be closely tied to $x_{\hi}$: It is the charge exchange equilibrium between $\oi$ and $\hi$ that provides the physical motivation for this work. As stated in Section~\ref{ssec:xHI_Omega_ratios}, $\Omega$ for both $\oi$ and $\si$ are calculated using Equation~\ref{eq:comoving_mass_density_absorbers}, using sums of the column densities of detected absorbers from our spectrum-generating code. This differs from the process utilized for $x_{\oi,\mathrm{LOS}}$ and $x_{\hi,\mathrm{LOS}}$ of summing \textit{all} the particles encountering the sightline segments. This modification means that the subset of particles selected have differing average characteristics compared to the distribution of all particles. For example, we find that particles contributing to absorption systems in $\oi$ have higher typical mass-weighted metallicities. Although not shown here, we further find that the particles in absorption systems generally occupy a narrower range of physical conditions, for example, in metallicity and total hydrogen column density. Ultimately, considering only absorption systems in the calculation of $\Omega_{\oi\si}$ means we are probing gas with a smaller range of physical characteristics. This is the reason for the decreased scatter between $x_{\hi}$ and $\Omega_{\oi\si}$ compared to $x_{\oi, \mathrm{LOS}}$.

\begin{figure}
\centering
\includegraphics[width=0.5\textwidth]{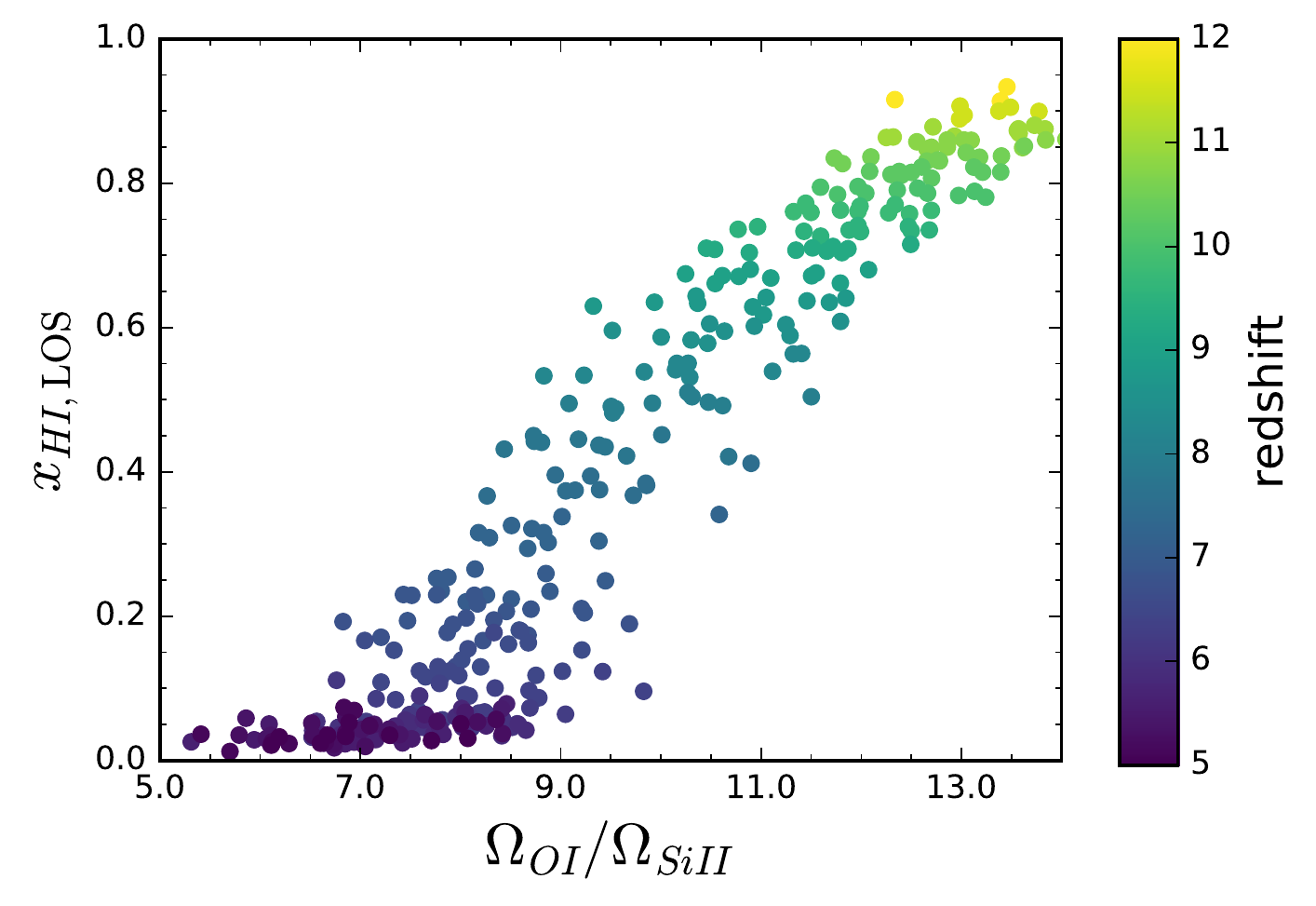}
\caption{The neutral hydrogen fraction measured from individual sightline segments as a function of the comoving density ratio of $\oi$ to $\siii$, where the colors indicate the redshift. The relationship is far less linear than when using $\Omega_{\oi\si}$ and shows increased scatter.}\label{fig:xHI_OmegaOISiII}
\end{figure}

The effect is illustrated further by a result from the investigation of~\citet{keating14}: they determined that $\oi$ could serve as a tracer of $\hi$ if it is ``well-shielded''. By directly computing the neutral oxygen fraction using {\sc cloudy}, assuming a~\citet{hm12} ionizing background, they find that the neutral oxygen fraction at $z=6$ can trace the neutral hydrogen fraction within 0.1 dex for regions with $\log N_{\hi} \geq 17.0$~\citep[Figure A1 of][]{keating14}. The self-shielding of such systems results in suppression of variability due to the amplitude of the UVB or the gas temperature, leading also to reduced variability in the ratio of the neutral oxygen fraction to the neutral hydrogen fraction. The consequences of this effect may be a secondary reason why we observe reduced scatter in the neutral hydrogen fraction's relationship with $\Omega_{\oi\si}$ compared to $x_{\oi}$, if absorbers are more likely to be dominated by well-shielded gas.

It is unclear if $\oi$, $\siii$, and $\siiv$ are likely to all be observed at reionization-era redshifts since some studies have shown that high-ionization states, such as $\siiv$, become less common as redshift increases~\citep{becker11,cooper19}. Indeed,~\citet{becker11} was only able to establish upper limits on the $\siiv$ column densities in their observed systems. Thus, if $\siii$ is the only ionization state of silicon that is likely to be observed for $z>6$, the effectiveness of $\Omega_{\oi\si}$ metric may be altered. We re-generate Figure~\ref{fig:xHI-proxies} excluding the comoving density in $\siiv$ to determine if the method is still useful in the absence of $\siiv$ measurements  in Figure~\ref{fig:xHI_OmegaOISiII}.

For $\Omega_{\oi\siii}$, similar to $\Omega_{\oi\si}$, there are two apparent relationships distinct on opposing sides of $x_{\hi}\approx 0.1$. For $\Omega_{\oi\siii} \lessapprox 7.0$, the relationship between this ratio and the neutral hydrogen fraction is nearly a flat line, indicating little to no relationship between the values. For high $\Omega_{\oi\siii}$ values ($\geq 9.0$) where the total comoving density in silicon is dominated by $\siii$, the highly linear relationship is maintained. However, the relationship is non-linear for lower $\Omega_{\oi\siii}$, and shows a larger degree of scatter when compared to $\Omega_{\oi\si}$. For these reasons, this metric ultimately does not appear to be as sensitive to $x_{\hi, \mathrm{LOS}}$ as $\Omega_{\oi\si}$.

\subsection{Caveats}
A significant item of consideration for our results is whether they are truly representative of the reionization era. Previous comparison to observations has shown that these simulations produce a plausible reionization history~\citep[Figure 8 of][]{finlator18}. To directly compare some values, \citet{ouchi18} used survey data from $6 \leq z \leq 7$ to identify LAEs and characterize their clustering properties. They found that the inherent clustering of LAEs is compounded by the patchiness of reionization and, in tandem with theoretical models, predicted the progression of reionization with redshift, concluding that $x_{\hi} = 0.15^{+0.15}_{-0.15}$ at $z=6.6$. Through analysis of the dispersion of $\tau_{\alpha}$, the Ly$\alpha$ optical depth,~\citet{inoue18} obtained a neutral fraction of $x_{\hi}=0.5^{+0.1}_{-0.3}$ at $z=7.3$. In this work, our simulated volume-averaged neutral fraction reaches $x_{\hi,v}=0.5$ at $z=8.2$ and $x_{\hi,v}=0.15$ at $z=6.9$, indicating a more gradual transition occurring earlier in the universe's history than is suggested by these results, finishing up to $\Delta z \sim 0.5$ early. However, our results do fall within the error bars established in these studies.

The simulation also predicts the large jump in $\oi$ incidence rate to occur at $z\approx6.3$, while the significant decrease in the observations of~\citet{becker19} occurs across $z=5.7$. As with the comparison of the inferred and simulated $x_{\hi, v}$, this apparent difference in timing may be a byproduct of the potentially early reionization history occurring in the simulation. We don't anticipate that the general trends would be severely affected by a change in the timing of the reionization history, since they are based on the physical conditions in the gas. It is possible, however, that the strong evolution we observe in the simulations from $z=7\rightarrow6$ of the column density distribution, equivalent width distribution, and covering fractions of absorbers may occur at a different point in cosmic time than we predict.

When compared to~\citet{becker19}, the simulation output at $z=6$ and 5 generates an equivalent width distribution in $\oi$ that overproduces the number density of very weak absorbers ($\mathrm{EW} < 0.03 \angstrom$) and underproduces stronger ones ($\mathrm{EW} \geq 0.1 \angstrom$), even after applying a completeness correction to the data. If the simulations are indeed overproducing low EW $\oi$ systems, it is possible that the differential change in incidence rate with redshift might be less obvious than the simulations indicate, since restricting our systems to $\mathrm{EW} \geq 0.05 \angstrom$ greatly diminishes the amplitude of the evolution. However, our incidence rates are consistent with the observed values, and although the extent of the evolution in $\oi$ is reduced when we include a $0.05 \angstrom$ cutoff, the decrease is still present and should be detectable given sufficient observational pathlength. Further, the results of~\citet{becker19} show a large change in incidence rate even for systems of $\mathrm{EW} \geq 0.05 \angstrom$, indicating that the evolution is not restricted to the weakest absorbers.

It may be that the overproduction of $z\approx5$ $\oi$ systems in the simulation is related to the strength of the UVB.~\citet{finlator18} demonstrated that the simulation's mean transmission at the Ly-$\alpha$ edge, $\tau_{\mathrm{Ly}\alpha}$, is in agreement with observations at $z\sim6$, but by $z\sim5$ has become too low by roughly a factor of 2 ~\citep{bosman18}. $\tau_{\mathrm{Ly}\alpha}$ is directly affected by the strength of the UVB, so it is plausible that the simulated UVB is not strong enough at $z\sim5$, leading to an unrealistically unaffected population of $\oi$ systems. This could reasonably explain why the simulated EW distribution and incidence rates show less evolution between $z=6$ and 5 than is suggested by~\citet{becker19}.

Examining the EW distribution of the $z_{\oi}\geq4.9$ observations in~\citet{becker19}, there is increasing deviation for lower EW between the $z<5.7$ and $z>5.7$ bins, with no apparent change in the number of systems with $\mathrm{EW} \geq 0.3$. This supports the qualitative trend we observe in the simulation that shows larger decreases in number for lower EW with decreasing redshift. However, even with completeness corrections, the simulation significantly overproduces weak absorption systems and underproduces strong ones, although it successfully matches the total number of systems with $\mathrm{EW} \geq 0.05 \angstrom$. This may be a side effect of resolution limitations within the simulation, as recent investigations have shown that implementing very high spatial resolutions ($\lessapprox 1$ kpc) in the CGM creates a larger number of low ionization systems such as $\oi$ and $\hi$, which would necessarily increase $\ell\left(X\right)$~\citep{hummels18,peeples19,vandevoort19}. Our more modest spatial resolution of tens of $\hckpc$ may be insufficient to capture the full detail of low ionization cloud structures. However, increased resolution also tends to increase the number of low-EW systems, so it is unclear whether this would improve or worsen the agreement in the EW distribution.

\citet{becker19} integrate the EW values of their low ionization states over a range of velocities, informed by their analysis of the line profiles. While we noted in Section~\ref{ssec:simulated_absorbers} that a merger velocity of 200 km/s doesn't significantly affect our column density distribution of absorption systems, in some cases they integrated over velocities up to 670 km/s. To test the dependence of this much larger velocity range, we examined our EW distributions using a merging velocity of 700 km/s. This adjustment does somewhat decrease the discrepancy between the observations and simulations, more so for smaller EW. At $z=6$, the simulated curve for $\mathrm{EW} \geq 0.03 \angstrom$, using this larger merging velocity would perfectly overlap the observations, but for $\mathrm{EW} \geq 0.1 \angstrom$ we find the simulated curve still falls markedly short of the observed values. Although the $z=5$ simulated curve also becomes flatter and closer to the observations using a higher merging velocity, there is no additional overlap in the curves. Taking these results into account, the disparity between our simulated EW distributions and~\citet{becker19} is likely not a byproduct of different merging criteria.

\section{Conclusions}\label{sec:conclusions}
Using these new simulations, we are able to make predictions that better represent the effects of a spatially variable ionizing background during reionization, and find that the evolution of the UVB in this time period is reflected in the abundance of neutral oxygen absorption systems. For the evolution of $\oi$, the general trends during the $z=8\rightarrow5$ EoR are as follows:
\begin{itemize}
\item The cross-section of neutral oxygen around halos is shrinking, even as the total oxygen distribution around halos is concurrently increasing (Figure~\ref{fig:OI_O_coldens_map});
\item The covering fraction of $\oi$ decreases around halos (Figure~\ref{fig:covering_fraction});
\item Weak $\oi$ absorbers are more affected than strong ones by the evolving UVB, showing a larger decrease in incidence at the tail end of reionization (Figure~\ref{fig:EW_OI});
\item $\oi$ absorbers far from halos and in less overdense gas are preferentially reduced compared to those that are nearby and in denser gas (Figures~\ref{fig:changing_distance_withz} and ~\ref{fig:OI_mass_fraction});
\item The simulated incidence rate of $\oi$ systems drops abruptly when $x_{\hi, v}$ reaches $\approx1$ percent, concurrent with strong evolution in the volume-averaged photoionization rate in the simulation. The change is stronger for weaker systems (Figure~\ref{fig:incidence_rate_xHIv}); 
\item There is a linear relationship between a proxy for the oxygen neutral fraction, $\Omega_{\oi\si}$, and the neutral hydrogen fraction in metal absorption systems, $x_{\hi, \mathrm{abs}}$. The relationship deviates from linearity once gas outside absorption systems is considered, but the scatter in the relationship almost invariably remains $<10$ percent (Figure~\ref{fig:xHI-proxies}).
\end{itemize}

Implementation of an on-the-fly radiative transfer in a hydrodynamic cosmological simulation shows that, concurrent with the epoch of overlap phase of hydrogen reionization, $\oi$ absorbers undergo a decrease in their footprint around halos and a decrease in the number of observed systems. This decrease occurs preferentially for absorbers that are farther from halos and that are weaker in terms of their equivalent width. Technicolor Dawn almost perfectly reproduces the number of observed $\oi$ systems with $\mathrm{EW} \geq 0.05 \angstrom$ at $z=6$, but underpredicts the number of systems at somewhat higher equivalent widths. By $z=5$, the simulation substantially overpredicts the number of $\mathrm{EW} \geq 0.05 \angstrom$ $\oi$ systems, perhaps indicating that such systems are not affected by the evolving UVB as strongly in the simulation as they are in reality. This may be the result of a too-weak simulated UVB at $z=5$, which is supported by the low Ly-$\alpha$ transmission reported in~\citet{finlator18}. Nonetheless, there is qualitative agreement that a sudden change in the incidence of $\oi$ is associated with the completion of $\hi$ reionization, and that evolution may be stronger among weaker absorbers.

The qualitative trends from both the simulation and observations are all consistent with an ``outside-in'' reionization scenario in which material occupying low overdensities is the first to feel the effects of a hardening UVB due to the lack of self-shielding in low density gas. Further, the consistent relationship between the neutral oxygen and neutral hydrogen fractions indicates that observations of neutral oxygen absorption systems could be used as a proxy to determine the mass-weighted neutral hydrogen fraction and thus the progression of reionization.

\section*{Acknowledgements}
The authors thank Chris Churchill for helpful discussions, comments, and assistance in implementing Voigt profile fits. The authors further thank George Becker and Laura Keating for inspiring conversations. Our simulation was run on COMET, part of the Extreme Science and Engineering Discovery Environment (XSEDE), which is supported by National Science Foundation grant number ACI-1548562. CCD thanks the LSSTC Data Science Fellowship Program, which is funded by LSSTC, NSF Cybertraining Grant \#1829740, the Brinson Foundation, and the Moore Foundation; Her participation in the program has benefited this work. CCD acknowledges funding from the New Mexico Space Grant Consortium Grant \#NNX15AL51H. Our work made use of the WebPlotDigitizer tool (https://automeris.io/WebPlotDigitizer),
for which we thank A. Rohatgi.



\bibliographystyle{mnras}
\bibliography{paper.bib}

\begin{thebibliography}{}
\makeatletter
\relax
\def\mn@urlcharsother{\let\do\@makeother \do\$\do\&\do\#\do\^\do\_\do\%\do\~}
\def\mn@doi{\begingroup\mn@urlcharsother \@ifnextchar [ {\mn@doi@}
  {\mn@doi@[]}}
\def\mn@doi@[#1]#2{\def\@tempa{#1}\ifx\@tempa\@empty \href
  {http://dx.doi.org/#2} {doi:#2}\else \href {http://dx.doi.org/#2} {#1}\fi
  \endgroup}
\def\mn@eprint#1#2{\mn@eprint@#1:#2::\@nil}
\def\mn@eprint@arXiv#1{\href {http://arxiv.org/abs/#1} {{\tt arXiv:#1}}}
\def\mn@eprint@dblp#1{\href {http://dblp.uni-trier.de/rec/bibtex/#1.xml}
  {dblp:#1}}
\def\mn@eprint@#1:#2:#3:#4\@nil{\def\@tempa {#1}\def\@tempb {#2}\def\@tempc
  {#3}\ifx \@tempc \@empty \let \@tempc \@tempb \let \@tempb \@tempa \fi \ifx
  \@tempb \@empty \def\@tempb {arXiv}\fi \@ifundefined
  {mn@eprint@\@tempb}{\@tempb:\@tempc}{\expandafter \expandafter \csname
  mn@eprint@\@tempb\endcsname \expandafter{\@tempc}}}

\bibitem[\protect\citeauthoryear{{Ba{\~n}ados} et~al.,}{{Ba{\~n}ados}
  et~al.}{2016}]{banados16}
{Ba{\~n}ados} E.,  et~al., 2016, \mn@doi [\apjs] {10.3847/0067-0049/227/1/11},
  \href {http://adsabs.harvard.edu/abs/2016ApJS..227...11B} {227, 11}

\bibitem[\protect\citeauthoryear{{Bahcall} \& {Peebles}}{{Bahcall} \&
  {Peebles}}{1969}]{bahcall69}
{Bahcall} J.~N.,  {Peebles} P.~J.~E.,  1969, \mn@doi [\apjl] {10.1086/180337},
  \href {http://adsabs.harvard.edu/abs/1969ApJ...156L...7B} {156, L7}

\bibitem[\protect\citeauthoryear{{Becker}, {Sargent}, {Rauch}  \&
  {Calverley}}{{Becker} et~al.}{2011}]{becker11}
{Becker} G.~D.,  {Sargent} W.~L.~W.,  {Rauch} M.,   {Calverley} A.~P.,  2011,
  \mn@doi [\apj] {10.1088/0004-637X/735/2/93}, \href
  {http://adsabs.harvard.edu/abs/2011ApJ...735...93B} {735, 93}

\bibitem[\protect\citeauthoryear{{Becker}, {Bolton}, {Madau}, {Pettini},
  {Ryan-Weber}  \& {Venemans}}{{Becker} et~al.}{2015}]{becker15mnras}
{Becker} G.~D.,  {Bolton} J.~S.,  {Madau} P.,  {Pettini} M.,  {Ryan-Weber}
  E.~V.,   {Venemans} B.~P.,  2015, \mn@doi [\mnras] {10.1093/mnras/stu2646},
  \href {http://adsabs.harvard.edu/abs/2015MNRAS.447.3402B} {447, 3402}

\bibitem[\protect\citeauthoryear{{Becker} et~al.,}{{Becker}
  et~al.}{2019}]{becker19}
{Becker} G.~D.,  et~al., 2019, arXiv e-prints, \href
  {https://ui.adsabs.harvard.edu/abs/2019arXiv190702983B} {p. arXiv:1907.02983}

\bibitem[\protect\citeauthoryear{{Bolton} \& {Haehnelt}}{{Bolton} \&
  {Haehnelt}}{2013}]{bolton13}
{Bolton} J.~S.,  {Haehnelt} M.~G.,  2013, \mn@doi [\mnras]
  {10.1093/mnras/sts455}, \href
  {http://adsabs.harvard.edu/abs/2013MNRAS.429.1695B} {429, 1695}

\bibitem[\protect\citeauthoryear{{Bosman}, {Fan}, {Jiang}, {Reed}, {Matsuoka},
  {Becker}  \& {Haehnelt}}{{Bosman} et~al.}{2018}]{bosman18}
{Bosman} S. E.~I.,  {Fan} X.,  {Jiang} L.,  {Reed} S.,  {Matsuoka} Y.,
  {Becker} G.,   {Haehnelt} M.,  2018, \mn@doi [\mnras]
  {10.1093/mnras/sty1344}, \href
  {https://ui.adsabs.harvard.edu/abs/2018MNRAS.479.1055B} {479, 1055}

\bibitem[\protect\citeauthoryear{{Bowman}, {Rogers}, {Monsalve}, {Mozdzen}  \&
  {Mahesh}}{{Bowman} et~al.}{2018}]{bowman18}
{Bowman} J.~D.,  {Rogers} A.~E.~E.,  {Monsalve} R.~A.,  {Mozdzen} T.~J.,
  {Mahesh} N.,  2018, \mn@doi [\nat] {10.1038/nature25792}, \href
  {http://adsabs.harvard.edu/abs/2018Natur.555...67B} {555, 67}

\bibitem[\protect\citeauthoryear{{Cai}, {Fan}, {Dave}, {Finlator}  \&
  {Oppenheimer}}{{Cai} et~al.}{2017}]{cai17}
{Cai} Z.,  {Fan} X.,  {Dave} R.,  {Finlator} K.,   {Oppenheimer} B.,  2017,
  \mn@doi [\apjl] {10.3847/2041-8213/aa8fc6}, \href
  {http://adsabs.harvard.edu/abs/2017ApJ...849L..18C} {849, L18}

\bibitem[\protect\citeauthoryear{{Chardin}, {Haehnelt}, {Aubert}  \&
  {Puchwein}}{{Chardin} et~al.}{2015}]{chardin15}
{Chardin} J.,  {Haehnelt} M.~G.,  {Aubert} D.,   {Puchwein} E.,  2015, \mn@doi
  [\mnras] {10.1093/mnras/stv1786}, \href
  {http://adsabs.harvard.edu/abs/2015MNRAS.453.2943C} {453, 2943}

\bibitem[\protect\citeauthoryear{{Chardin}, {Puchwein}  \&
  {Haehnelt}}{{Chardin} et~al.}{2017}]{chardin17}
{Chardin} J.,  {Puchwein} E.,   {Haehnelt} M.~G.,  2017, \mn@doi [\mnras]
  {10.1093/mnras/stw2943}, \href
  {http://adsabs.harvard.edu/abs/2017MNRAS.465.3429C} {465, 3429}

\bibitem[\protect\citeauthoryear{{Codoreanu}, {Ryan-Weber}, {Garc{\'{\i}}a},
  {Crighton}, {Becker}, {Pettini}, {Madau}  \& {Venemans}}{{Codoreanu}
  et~al.}{2018}]{codoreanu18}
{Codoreanu} A.,  {Ryan-Weber} E.~V.,  {Garc{\'{\i}}a} L.~{\'A}.,  {Crighton}
  N.~H.~M.,  {Becker} G.,  {Pettini} M.,  {Madau} P.,   {Venemans} B.,  2018,
  \mn@doi [\mnras] {10.1093/mnras/sty2576}, \href
  {http://adsabs.harvard.edu/abs/2018MNRAS.481.4940C} {481, 4940}

\bibitem[\protect\citeauthoryear{{Cooper}, {Simcoe}, {Cooksey}, {Bordoloi},
  {Miller}, {Furezs}, {Turner}  \& {Ba{\~n}ados}}{{Cooper}
  et~al.}{2019}]{cooper19}
{Cooper} T.~J.,  {Simcoe} R.~A.,  {Cooksey} K.~L.,  {Bordoloi} R.,  {Miller}
  D.~R.,  {Furezs} G.,  {Turner} M.~L.,   {Ba{\~n}ados} E.,  2019, arXiv
  e-prints, \href {http://adsabs.harvard.edu/abs/2019arXiv190105980C} {}

\bibitem[\protect\citeauthoryear{{Couchman} \& {Rees}}{{Couchman} \&
  {Rees}}{1986}]{couchman86}
{Couchman} H.~M.~P.,  {Rees} M.~J.,  1986, \mn@doi [\mnras]
  {10.1093/mnras/221.1.53}, \href
  {http://adsabs.harvard.edu/abs/1986MNRAS.221...53C} {221, 53}

\bibitem[\protect\citeauthoryear{{D'Aloisio}, {Upton Sanderbeck}, {McQuinn},
  {Trac}  \& {Shapiro}}{{D'Aloisio} et~al.}{2017}]{daloisio17}
{D'Aloisio} A.,  {Upton Sanderbeck} P.~R.,  {McQuinn} M.,  {Trac} H.,
  {Shapiro} P.~R.,  2017, \mn@doi [\mnras] {10.1093/mnras/stx711}, \href
  {http://adsabs.harvard.edu/abs/2017MNRAS.468.4691D} {468, 4691}

\bibitem[\protect\citeauthoryear{{Dav{\'e}}, {Thompson}  \&
  {Hopkins}}{{Dav{\'e}} et~al.}{2016}]{dave16}
{Dav{\'e}} R.,  {Thompson} R.,   {Hopkins} P.~F.,  2016, \mn@doi [\mnras]
  {10.1093/mnras/stw1862}, \href
  {http://adsabs.harvard.edu/abs/2016MNRAS.462.3265D} {462, 3265}

\bibitem[\protect\citeauthoryear{{DeBoer}}{{DeBoer}}{2016}]{deboer16}
{DeBoer} D.~R.,  2016, in Electromagnetics in Advanced Applications (ICEAA),
  2016 International Conference on, article id. 7731446. p. 7731446,
  \mn@doi{10.1109/ICEAA.2016.7731446}

\bibitem[\protect\citeauthoryear{{Doughty}, {Finlator}, {Oppenheimer},
  {Dav{\'e}}  \& {Zackrisson}}{{Doughty} et~al.}{2018}]{doughty18}
{Doughty} C.,  {Finlator} K.,  {Oppenheimer} B.~D.,  {Dav{\'e}} R.,
  {Zackrisson} E.,  2018, \mn@doi [\mnras] {10.1093/mnras/sty156}, \href
  {http://adsabs.harvard.edu/abs/2018MNRAS.tmp..152D} {}

\bibitem[\protect\citeauthoryear{{Einasto} \& {Einasto}}{{Einasto} \&
  {Einasto}}{1987}]{einasto87}
{Einasto} M.,  {Einasto} J.,  1987, \mn@doi [\mnras] {10.1093/mnras/226.3.543},
  \href {http://adsabs.harvard.edu/abs/1987MNRAS.226..543E} {226, 543}

\bibitem[\protect\citeauthoryear{{Fan}, {Narayanan}, {Strauss}, {White},
  {Becker}, {Pentericci}  \& {Rix}}{{Fan} et~al.}{2002}]{fan02}
{Fan} X.,  {Narayanan} V.~K.,  {Strauss} M.~A.,  {White} R.~L.,  {Becker}
  R.~H.,  {Pentericci} L.,   {Rix} H.-W.,  2002, \mn@doi [\aj]
  {10.1086/339030}, \href {http://adsabs.harvard.edu/abs/2002AJ....123.1247F}
  {123, 1247}

\bibitem[\protect\citeauthoryear{{Fan}, {Carilli}  \& {Keating}}{{Fan}
  et~al.}{2006}]{fan06}
{Fan} X.,  {Carilli} C.~L.,   {Keating} B.,  2006, \mn@doi [\araa]
  {10.1146/annurev.astro.44.051905.092514}, \href
  {http://adsabs.harvard.edu/abs/2006ARA%26A..44..415F} {44, 415}

\bibitem[\protect\citeauthoryear{{Ferland}, {Korista}, {Verner}, {Ferguson},
  {Kingdon}  \& {Verner}}{{Ferland} et~al.}{1998}]{ferland98}
{Ferland} G.~J.,  {Korista} K.~T.,  {Verner} D.~A.,  {Ferguson} J.~W.,
  {Kingdon} J.~B.,   {Verner} E.~M.,  1998, \mn@doi [\pasp] {10.1086/316190},
  \href {http://adsabs.harvard.edu/abs/1998PASP..110..761F} {110, 761}

\bibitem[\protect\citeauthoryear{{Finlator}, {{\"O}zel}  \&
  {Dav{\'e}}}{{Finlator} et~al.}{2009}]{finlator09}
{Finlator} K.,  {{\"O}zel} F.,   {Dav{\'e}} R.,  2009, \mn@doi [\mnras]
  {10.1111/j.1365-2966.2008.14190.x}, \href
  {https://ui.adsabs.harvard.edu/abs/2009MNRAS.393.1090F} {393, 1090}

\bibitem[\protect\citeauthoryear{{Finlator}, {Mu{\~n}oz}, {Oppenheimer}, {Oh},
  {{\"O}zel}  \& {Dav{\'e}}}{{Finlator} et~al.}{2013}]{finlator13}
{Finlator} K.,  {Mu{\~n}oz} J.~A.,  {Oppenheimer} B.~D.,  {Oh} S.~P.,
  {{\"O}zel} F.,   {Dav{\'e}} R.,  2013, \mn@doi [\mnras]
  {10.1093/mnras/stt1697}, \href
  {http://adsabs.harvard.edu/abs/2013MNRAS.436.1818F} {436, 1818}

\bibitem[\protect\citeauthoryear{{Finlator}, {Thompson}, {Huang}, {Dav{\'e}},
  {Zackrisson}  \& {Oppenheimer}}{{Finlator} et~al.}{2015}]{finlator15}
{Finlator} K.,  {Thompson} R.,  {Huang} S.,  {Dav{\'e}} R.,  {Zackrisson} E.,
  {Oppenheimer} B.~D.,  2015, \mn@doi [\mnras] {10.1093/mnras/stu2668}, \href
  {http://adsabs.harvard.edu/abs/2015MNRAS.447.2526F} {447, 2526}

\bibitem[\protect\citeauthoryear{{Finlator}, {Keating}, {Oppenheimer},
  {Dav{\'e}}  \& {Zackrisson}}{{Finlator} et~al.}{2018}]{finlator18}
{Finlator} K.,  {Keating} L.,  {Oppenheimer} B.~D.,  {Dav{\'e}} R.,
  {Zackrisson} E.,  2018, \mn@doi [\mnras] {10.1093/mnras/sty1949}, \href
  {http://adsabs.harvard.edu/abs/2018MNRAS.480.2628F} {480, 2628}

\bibitem[\protect\citeauthoryear{{Fumagalli}, {O'Meara}, {Prochaska}  \&
  {Worseck}}{{Fumagalli} et~al.}{2013}]{fumagalli13}
{Fumagalli} M.,  {O'Meara} J.~M.,  {Prochaska} J.~X.,   {Worseck} G.,  2013,
  \mn@doi [\apj] {10.1088/0004-637X/775/1/78}, \href
  {http://adsabs.harvard.edu/abs/2013ApJ...775...78F} {775, 78}

\bibitem[\protect\citeauthoryear{{Garc{\'{\i}}a}, {Tescari}, {Ryan-Weber}  \&
  {Wyithe}}{{Garc{\'{\i}}a} et~al.}{2017}]{garcia17}
{Garc{\'{\i}}a} L.~A.,  {Tescari} E.,  {Ryan-Weber} E.~V.,   {Wyithe} J.~S.~B.,
   2017, \mn@doi [\mnras] {10.1093/mnrasl/slx053}, \href
  {http://adsabs.harvard.edu/abs/2017MNRAS.469L..53G} {469, L53}

\bibitem[\protect\citeauthoryear{{Glazer}, {Rau}  \& {Trac}}{{Glazer}
  et~al.}{2018}]{glazer18}
{Glazer} D.,  {Rau} M.~M.,   {Trac} H.,  2018, \mn@doi [Research Notes of the
  American Astronomical Society] {10.3847/2515-5172/aad68a}, \href
  {http://adsabs.harvard.edu/abs/2018RNAAS...2c.135G} {2, 135}

\bibitem[\protect\citeauthoryear{{Gnedin}}{{Gnedin}}{2000}]{gnedin00}
{Gnedin} N.~Y.,  2000, \mn@doi [\apj] {10.1086/308876}, \href
  {https://ui.adsabs.harvard.edu/abs/2000ApJ...535..530G} {535, 530}

\bibitem[\protect\citeauthoryear{{Gunn} \& {Peterson}}{{Gunn} \&
  {Peterson}}{1965}]{gunn65}
{Gunn} J.~E.,  {Peterson} B.~A.,  1965, \mn@doi [\apj] {10.1086/148444}, \href
  {http://adsabs.harvard.edu/abs/1965ApJ...142.1633G} {142, 1633}

\bibitem[\protect\citeauthoryear{{Haardt} \& {Madau}}{{Haardt} \&
  {Madau}}{2012}]{hm12}
{Haardt} F.,  {Madau} P.,  2012, \mn@doi [\apj] {10.1088/0004-637X/746/2/125},
  \href {http://adsabs.harvard.edu/abs/2012ApJ...746..125H} {746, 125}

\bibitem[\protect\citeauthoryear{{Hahn} \& {Abel}}{{Hahn} \&
  {Abel}}{2011}]{hahn11}
{Hahn} O.,  {Abel} T.,  2011, \mn@doi [\mnras]
  {10.1111/j.1365-2966.2011.18820.x}, \href
  {http://adsabs.harvard.edu/abs/2011MNRAS.415.2101H} {415, 2101}

\bibitem[\protect\citeauthoryear{{Haiman} \& {Loeb}}{{Haiman} \&
  {Loeb}}{1997}]{haiman97}
{Haiman} Z.,  {Loeb} A.,  1997, \mn@doi [\apj] {10.1086/304238}, \href
  {http://adsabs.harvard.edu/abs/1997ApJ...483...21H} {483, 21}

\bibitem[\protect\citeauthoryear{{Heringer}, {Pritchet}, {Kezwer}, {Graham},
  {Sand}  \& {Bildfell}}{{Heringer} et~al.}{2017}]{heringer17}
{Heringer} E.,  {Pritchet} C.,  {Kezwer} J.,  {Graham} M.~L.,  {Sand} D.,
  {Bildfell} C.,  2017, \mn@doi [\apj] {10.3847/1538-4357/834/1/15}, \href
  {http://adsabs.harvard.edu/abs/2017ApJ...834...15H} {834, 15}

\bibitem[\protect\citeauthoryear{{Hoag} et~al.,}{{Hoag} et~al.}{2019}]{hoag19}
{Hoag} A.,  et~al., 2019, arXiv e-prints, \href
  {http://adsabs.harvard.edu/abs/2019arXiv190109001H} {}

\bibitem[\protect\citeauthoryear{{Huml{\'{\i}}cek}}{{Huml{\'{\i}}cek}}{1979}]{humlicek79}
{Huml{\'{\i}}cek} J.,  1979, \mn@doi [\jqsrt] {10.1016/0022-4073(79)90062-1},
  \href {http://adsabs.harvard.edu/abs/1979JQSRT..21..309H} {21, 309}

\bibitem[\protect\citeauthoryear{{Hummels} et~al.,}{{Hummels}
  et~al.}{2018}]{hummels18}
{Hummels} C.~B.,  et~al., 2018, arXiv e-prints, \href
  {https://ui.adsabs.harvard.edu/abs/2018arXiv181112410H} {p. arXiv:1811.12410}

\bibitem[\protect\citeauthoryear{{Inoue} et~al.,}{{Inoue}
  et~al.}{2018}]{inoue18}
{Inoue} A.~K.,  et~al., 2018, \mn@doi [\pasj] {10.1093/pasj/psy048}, \href
  {http://adsabs.harvard.edu/abs/2018PASJ...70...55I} {70, 55}

\bibitem[\protect\citeauthoryear{{Kacprzak} et~al.,}{{Kacprzak}
  et~al.}{2019}]{kacprzak19}
{Kacprzak} G.~G.,  et~al., 2019, \mn@doi [\apj] {10.3847/1538-4357/aaf1a6},
  \href {http://adsabs.harvard.edu/abs/2019ApJ...870..137K} {870, 137}

\bibitem[\protect\citeauthoryear{{Keating}, {Haehnelt}, {Becker}  \&
  {Bolton}}{{Keating} et~al.}{2014}]{keating14}
{Keating} L.~C.,  {Haehnelt} M.~G.,  {Becker} G.~D.,   {Bolton} J.~S.,  2014,
  \mn@doi [\mnras] {10.1093/mnras/stt2324}, \href
  {http://adsabs.harvard.edu/abs/2014MNRAS.438.1820K} {438, 1820}

\bibitem[\protect\citeauthoryear{{Keating}, {Puchwein}, {Haehnelt}, {Bird}  \&
  {Bolton}}{{Keating} et~al.}{2016}]{keating16}
{Keating} L.~C.,  {Puchwein} E.,  {Haehnelt} M.~G.,  {Bird} S.,   {Bolton}
  J.~S.,  2016, \mn@doi [\mnras] {10.1093/mnras/stw1306}, \href
  {http://adsabs.harvard.edu/abs/2016MNRAS.461..606K} {461, 606}

\bibitem[\protect\citeauthoryear{{Konno} et~al.,}{{Konno}
  et~al.}{2018}]{konno18}
{Konno} A.,  et~al., 2018, \mn@doi [\pasj] {10.1093/pasj/psx131}, \href
  {http://adsabs.harvard.edu/abs/2018PASJ...70S..16K} {70, S16}

\bibitem[\protect\citeauthoryear{{Kroupa}}{{Kroupa}}{2001}]{kroupa01}
{Kroupa} P.,  2001, \mn@doi [\mnras] {10.1046/j.1365-8711.2001.04022.x}, \href
  {http://adsabs.harvard.edu/abs/2001MNRAS.322..231K} {322, 231}

\bibitem[\protect\citeauthoryear{{Madau} \& {Rees}}{{Madau} \&
  {Rees}}{2000}]{madau00}
{Madau} P.,  {Rees} M.~J.,  2000, \mn@doi [\apjl] {10.1086/312934}, \href
  {http://adsabs.harvard.edu/abs/2000ApJ...542L..69M} {542, L69}

\bibitem[\protect\citeauthoryear{{Mason}, {Treu}, {Dijkstra}, {Mesinger},
  {Trenti}, {Pentericci}, {de Barros}  \& {Vanzella}}{{Mason}
  et~al.}{2018}]{mason18}
{Mason} C.~A.,  {Treu} T.,  {Dijkstra} M.,  {Mesinger} A.,  {Trenti} M.,
  {Pentericci} L.,  {de Barros} S.,   {Vanzella} E.,  2018, \mn@doi [\apj]
  {10.3847/1538-4357/aab0a7}, \href
  {http://adsabs.harvard.edu/abs/2018ApJ...856....2M} {856, 2}

\bibitem[\protect\citeauthoryear{{McGreer}, {Fan}, {Jiang}  \& {Cai}}{{McGreer}
  et~al.}{2018}]{mcgreer18}
{McGreer} I.~D.,  {Fan} X.,  {Jiang} L.,   {Cai} Z.,  2018, \mn@doi [\aj]
  {10.3847/1538-3881/aaaab4}, \href
  {http://adsabs.harvard.edu/abs/2018AJ....155..131M} {155, 131}

\bibitem[\protect\citeauthoryear{{McQuinn}, {Lidz}, {Zahn}, {Dutta},
  {Hernquist}  \& {Zaldarriaga}}{{McQuinn} et~al.}{2007}]{mcquinn07}
{McQuinn} M.,  {Lidz} A.,  {Zahn} O.,  {Dutta} S.,  {Hernquist} L.,
  {Zaldarriaga} M.,  2007, \mn@doi [\mnras] {10.1111/j.1365-2966.2007.11489.x},
  \href {http://adsabs.harvard.edu/abs/2007MNRAS.377.1043M} {377, 1043}

\bibitem[\protect\citeauthoryear{{Mellema} et~al.,}{{Mellema}
  et~al.}{2013}]{mellema13}
{Mellema} G.,  et~al., 2013, \mn@doi [Experimental Astronomy]
  {10.1007/s10686-013-9334-5}, \href
  {http://adsabs.harvard.edu/abs/2013ExA....36..235M} {36, 235}

\bibitem[\protect\citeauthoryear{{Miralda-Escud{\'e}}}{{Miralda-Escud{\'e}}}{1998}]{miralda98}
{Miralda-Escud{\'e}} J.,  1998, \mn@doi [\apj] {10.1086/305799}, \href
  {http://adsabs.harvard.edu/abs/1998ApJ...501...15M} {501, 15}

\bibitem[\protect\citeauthoryear{{Miralda-Escud{\'e}}, {Haehnelt}  \&
  {Rees}}{{Miralda-Escud{\'e}} et~al.}{2000}]{miralda00}
{Miralda-Escud{\'e}} J.,  {Haehnelt} M.,   {Rees} M.~J.,  2000, \mn@doi [\apj]
  {10.1086/308330}, \href {http://adsabs.harvard.edu/abs/2000ApJ...530....1M}
  {530, 1}

\bibitem[\protect\citeauthoryear{{Morales} \& {Wyithe}}{{Morales} \&
  {Wyithe}}{2010}]{morales10}
{Morales} M.~F.,  {Wyithe} J.~S.~B.,  2010, \mn@doi [\araa]
  {10.1146/annurev-astro-081309-130936}, \href
  {http://adsabs.harvard.edu/abs/2010ARA%26A..48..127M} {48, 127}

\bibitem[\protect\citeauthoryear{{Mortlock}}{{Mortlock}}{2016}]{mortlock16}
{Mortlock} D.,  2016, in {Mesinger} A.,  ed.,  Astrophysics and Space Science
  Library Vol. 423, Understanding the Epoch of Cosmic Reionization: Challenges
  and Progress. p.~187 (\mn@eprint {arXiv} {1511.01107}),
  \mn@doi{10.1007/978-3-319-21957-8_7}

\bibitem[\protect\citeauthoryear{{Muratov}, {Kere{\v s}},
  {Faucher-Gigu{\`e}re}, {Hopkins}, {Quataert}  \& {Murray}}{{Muratov}
  et~al.}{2015}]{muratov15}
{Muratov} A.~L.,  {Kere{\v s}} D.,  {Faucher-Gigu{\`e}re} C.-A.,  {Hopkins}
  P.~F.,  {Quataert} E.,   {Murray} N.,  2015, \mn@doi [\mnras]
  {10.1093/mnras/stv2126}, \href
  {http://adsabs.harvard.edu/abs/2015MNRAS.454.2691M} {454, 2691}

\bibitem[\protect\citeauthoryear{{Nielsen}, {Kacprzak}, {Muzahid}, {Churchill},
  {Murphy}  \& {Charlton}}{{Nielsen} et~al.}{2017}]{nielsen17}
{Nielsen} N.~M.,  {Kacprzak} G.~G.,  {Muzahid} S.,  {Churchill} C.~W.,
  {Murphy} M.~T.,   {Charlton} J.~C.,  2017, \mn@doi [\apj]
  {10.3847/1538-4357/834/2/148}, \href
  {http://adsabs.harvard.edu/abs/2017ApJ...834..148N} {834, 148}

\bibitem[\protect\citeauthoryear{{Nomoto}, {Tominaga}, {Umeda}, {Kobayashi}  \&
  {Maeda}}{{Nomoto} et~al.}{2006}]{nomoto06}
{Nomoto} K.,  {Tominaga} N.,  {Umeda} H.,  {Kobayashi} C.,   {Maeda} K.,  2006,
  \mn@doi [Nuclear Physics A] {10.1016/j.nuclphysa.2006.05.008}, \href
  {http://adsabs.harvard.edu/abs/2006NuPhA.777..424N} {777, 424}

\bibitem[\protect\citeauthoryear{{Oh}}{{Oh}}{2002}]{oh02}
{Oh} S.~P.,  2002, \mn@doi [\mnras] {10.1046/j.1365-8711.2002.05859.x}, \href
  {http://adsabs.harvard.edu/abs/2002MNRAS.336.1021O} {336, 1021}

\bibitem[\protect\citeauthoryear{{Onoue} et~al.,}{{Onoue}
  et~al.}{2017}]{onoue17}
{Onoue} M.,  et~al., 2017, \mn@doi [\apjl] {10.3847/2041-8213/aa8cc6}, \href
  {http://adsabs.harvard.edu/abs/2017ApJ...847L..15O} {847, L15}

\bibitem[\protect\citeauthoryear{{Oppenheimer} \& {Dav{\'e}}}{{Oppenheimer} \&
  {Dav{\'e}}}{2008}]{oppenheimer08}
{Oppenheimer} B.~D.,  {Dav{\'e}} R.,  2008, \mn@doi [\mnras]
  {10.1111/j.1365-2966.2008.13280.x}, \href
  {https://ui.adsabs.harvard.edu/abs/2008MNRAS.387..577O} {387, 577}

\bibitem[\protect\citeauthoryear{{Oppenheimer}, {Dav{\'e}}  \&
  {Finlator}}{{Oppenheimer} et~al.}{2009}]{oppenheimer09}
{Oppenheimer} B.~D.,  {Dav{\'e}} R.,   {Finlator} K.,  2009, \mn@doi [\mnras]
  {10.1111/j.1365-2966.2009.14771.x}, \href
  {http://adsabs.harvard.edu/abs/2009MNRAS.396..729O} {396, 729}

\bibitem[\protect\citeauthoryear{{Oppenheimer} et~al.,}{{Oppenheimer}
  et~al.}{2016}]{oppenheimer16}
{Oppenheimer} B.~D.,  et~al., 2016, \mn@doi [\mnras] {10.1093/mnras/stw1066},
  \href {http://adsabs.harvard.edu/abs/2016MNRAS.460.2157O} {460, 2157}

\bibitem[\protect\citeauthoryear{{Oppenheimer}, {Schaye}, {Crain}, {Werk}  \&
  {Richings}}{{Oppenheimer} et~al.}{2018}]{oppenheimer18}
{Oppenheimer} B.~D.,  {Schaye} J.,  {Crain} R.~A.,  {Werk} J.~K.,   {Richings}
  A.~J.,  2018, \mn@doi [\mnras] {10.1093/mnras/sty2281}, \href
  {http://adsabs.harvard.edu/abs/2018MNRAS.481..835O} {481, 835}

\bibitem[\protect\citeauthoryear{{Ostriker} \& {Gnedin}}{{Ostriker} \&
  {Gnedin}}{1996}]{ostriker96}
{Ostriker} J.~P.,  {Gnedin} N.~Y.,  1996, \mn@doi [\apjl] {10.1086/310375},
  \href {http://adsabs.harvard.edu/abs/1996ApJ...472L..63O} {472, L63}

\bibitem[\protect\citeauthoryear{{Ouchi} et~al.,}{{Ouchi}
  et~al.}{2018}]{ouchi18}
{Ouchi} M.,  et~al., 2018, \mn@doi [\pasj] {10.1093/pasj/psx074}, \href
  {http://adsabs.harvard.edu/abs/2018PASJ...70S..13O} {70, S13}

\bibitem[\protect\citeauthoryear{{Pallottini}, {Gallerani}  \&
  {Ferrara}}{{Pallottini} et~al.}{2014}]{pallottini14}
{Pallottini} A.,  {Gallerani} S.,   {Ferrara} A.,  2014, \mn@doi [\mnras]
  {10.1093/mnrasl/slu126}, \href
  {http://adsabs.harvard.edu/abs/2014MNRAS.444L.105P} {444, L105}

\bibitem[\protect\citeauthoryear{{Patil} et~al.,}{{Patil}
  et~al.}{2017}]{patil17}
{Patil} A.~H.,  et~al., 2017, \mn@doi [\apj] {10.3847/1538-4357/aa63e7}, \href
  {http://adsabs.harvard.edu/abs/2017ApJ...838...65P} {838, 65}

\bibitem[\protect\citeauthoryear{{Peeples} et~al.,}{{Peeples}
  et~al.}{2019}]{peeples19}
{Peeples} M.~S.,  et~al., 2019, \mn@doi [\apj] {10.3847/1538-4357/ab0654},
  \href {https://ui.adsabs.harvard.edu/abs/2019ApJ...873..129P} {873, 129}

\bibitem[\protect\citeauthoryear{{Pehlivan Rhodin}, {Agertz}, {Christensen},
  {Renaud}  \& {Uldall Fynbo}}{{Pehlivan Rhodin} et~al.}{2019}]{rhodin19}
{Pehlivan Rhodin} N.~H.,  {Agertz} O.,  {Christensen} L.,  {Renaud} F.,
  {Uldall Fynbo} J.~P.,  2019, arXiv e-prints, \href
  {http://adsabs.harvard.edu/abs/2019arXiv190110777P} {}

\bibitem[\protect\citeauthoryear{{Planck Collaboration} et~al.,}{{Planck
  Collaboration} et~al.}{2016}]{planckcol16}
{Planck Collaboration} et~al., 2016, \mn@doi [\aap]
  {10.1051/0004-6361/201525830}, \href
  {http://adsabs.harvard.edu/abs/2016A%26A...594A..13P} {594, A13}

\bibitem[\protect\citeauthoryear{{Planck Collaboration} et~al.,}{{Planck
  Collaboration} et~al.}{2018}]{planckcol18}
{Planck Collaboration} et~al., 2018, preprint, \href
  {http://adsabs.harvard.edu/abs/2018arXiv180706209P} {} (\mn@eprint {arXiv}
  {1807.06209})

\bibitem[\protect\citeauthoryear{{Prochter}, {Prochaska}  \&
  {Burles}}{{Prochter} et~al.}{2006}]{prochter06}
{Prochter} G.~E.,  {Prochaska} J.~X.,   {Burles} S.~M.,  2006, \mn@doi [\apj]
  {10.1086/499341}, \href {http://adsabs.harvard.edu/abs/2006ApJ...639..766P}
  {639, 766}

\bibitem[\protect\citeauthoryear{{Rafelski}, {Wolfe}, {Prochaska}, {Neeleman}
  \& {Mendez}}{{Rafelski} et~al.}{2012}]{rafelski12}
{Rafelski} M.,  {Wolfe} A.~M.,  {Prochaska} J.~X.,  {Neeleman} M.,   {Mendez}
  A.~J.,  2012, \mn@doi [\apj] {10.1088/0004-637X/755/2/89}, \href
  {http://adsabs.harvard.edu/abs/2012ApJ...755...89R} {755, 89}

\bibitem[\protect\citeauthoryear{{Ryan-Weber}, {Pettini}, {Madau}  \&
  {Zych}}{{Ryan-Weber} et~al.}{2009}]{ryan-weber09}
{Ryan-Weber} E.~V.,  {Pettini} M.,  {Madau} P.,   {Zych} B.~J.,  2009, \mn@doi
  [\mnras] {10.1111/j.1365-2966.2009.14618.x}, \href
  {http://adsabs.harvard.edu/abs/2009MNRAS.395.1476R} {395, 1476}

\bibitem[\protect\citeauthoryear{{Seyffert}, {Cooksey}, {Simcoe}, {O'Meara},
  {Kao}  \& {Prochaska}}{{Seyffert} et~al.}{2013}]{seyffert13}
{Seyffert} E.~N.,  {Cooksey} K.~L.,  {Simcoe} R.~A.,  {O'Meara} J.~M.,  {Kao}
  M.~M.,   {Prochaska} J.~X.,  2013, \mn@doi [\apj]
  {10.1088/0004-637X/779/2/161}, \href
  {http://adsabs.harvard.edu/abs/2013ApJ...779..161S} {779, 161}

\bibitem[\protect\citeauthoryear{{Simcoe}, {Sullivan}, {Cooksey}, {Kao},
  {Matejek}  \& {Burgasser}}{{Simcoe} et~al.}{2012}]{simcoe12}
{Simcoe} R.~A.,  {Sullivan} P.~W.,  {Cooksey} K.~L.,  {Kao} M.~M.,  {Matejek}
  M.~S.,   {Burgasser} A.~J.,  2012, \mn@doi [\nat] {10.1038/nature11612},
  \href {http://adsabs.harvard.edu/abs/2012Natur.492...79S} {492, 79}

\bibitem[\protect\citeauthoryear{{Songaila}}{{Songaila}}{2001}]{songaila01}
{Songaila} A.,  2001, \mn@doi [\apjl] {10.1086/324761}, \href
  {http://adsabs.harvard.edu/abs/2001ApJ...561L.153S} {561, L153}

\bibitem[\protect\citeauthoryear{{Songaila} \& {Cowie}}{{Songaila} \&
  {Cowie}}{2010}]{songaila10}
{Songaila} A.,  {Cowie} L.~L.,  2010, \mn@doi [\apj]
  {10.1088/0004-637X/721/2/1448}, \href
  {http://adsabs.harvard.edu/abs/2010ApJ...721.1448S} {721, 1448}

\bibitem[\protect\citeauthoryear{{Springel} \& {Hernquist}}{{Springel} \&
  {Hernquist}}{2003}]{springel03}
{Springel} V.,  {Hernquist} L.,  2003, \mn@doi [\mnras]
  {10.1046/j.1365-8711.2003.06207.x}, \href
  {https://ui.adsabs.harvard.edu/abs/2003MNRAS.339..312S} {339, 312}

\bibitem[\protect\citeauthoryear{{Stark}}{{Stark}}{2016}]{stark16}
{Stark} D.~P.,  2016, \mn@doi [\araa] {10.1146/annurev-astro-081915-023417},
  \href {http://adsabs.harvard.edu/abs/2016ARA%26A..54..761S} {54, 761}

\bibitem[\protect\citeauthoryear{{Treu}, {Schmidt}, {Trenti}, {Bradley}  \&
  {Stiavelli}}{{Treu} et~al.}{2013}]{treu13}
{Treu} T.,  {Schmidt} K.~B.,  {Trenti} M.,  {Bradley} L.~D.,   {Stiavelli} M.,
  2013, \mn@doi [\apjl] {10.1088/2041-8205/775/1/L29}, \href
  {http://adsabs.harvard.edu/abs/2013ApJ...775L..29T} {775, L29}

\bibitem[\protect\citeauthoryear{{Tumlinson}, {Peeples}  \& {Werk}}{{Tumlinson}
  et~al.}{2017}]{tumlinson17}
{Tumlinson} J.,  {Peeples} M.~S.,   {Werk} J.~K.,  2017, \mn@doi [\araa]
  {10.1146/annurev-astro-091916-055240}, \href
  {http://adsabs.harvard.edu/abs/2017ARA%26A..55..389T} {55, 389}

\bibitem[\protect\citeauthoryear{{van de Voort}, {Springel}, {Mandelker}, {van
  den Bosch}  \& {Pakmor}}{{van de Voort} et~al.}{2019}]{vandevoort19}
{van de Voort} F.,  {Springel} V.,  {Mandelker} N.,  {van den Bosch} F.~C.,
  {Pakmor} R.,  2019, \mn@doi [\mnras] {10.1093/mnrasl/sly190}, \href
  {https://ui.adsabs.harvard.edu/abs/2019MNRAS.482L..85V} {482, L85}

\makeatother
\end{thebibliography}






\bsp	
\label{lastpage}
\end{document}